\documentclass{pasj01}

\begin{document}

\author{Mariko \textsc{Kimura}\altaffilmark{1,*},
        Taichi \textsc{Kato}\altaffilmark{1},
        Hiroyuki \textsc{Maehara}\altaffilmark{2,3},
        Ryoko \textsc{Ishioka}\altaffilmark{4,5},
        Berto \textsc{Monard}\altaffilmark{6,7},
        Kazuhiro \textsc{Nakajima}\altaffilmark{8},
        Geoff \textsc{Stone}\altaffilmark{9},
        Elena P. \textsc{Pavlenko}\altaffilmark{10},
        Oksana I. \textsc{Antonyuk}\altaffilmark{10},
        Nikolai V. \textsc{Pit}\altaffilmark{10},
        Aleksei A. \textsc{Sosnovskij}\altaffilmark{10},
        Natalia \textsc{Katysheva}\altaffilmark{11}, 
        Michael \textsc{Richmond}\altaffilmark{12},
        Ra\'{u}l \textsc{Michel}\altaffilmark{13}, 
        Katsura \textsc{Matsumoto}\altaffilmark{14},
        Naoto \textsc{Kojiguchi}\altaffilmark{14},
        Yuki \textsc{Sugiura}\altaffilmark{14},
        Shihei \textsc{Tei}\altaffilmark{14},
        Kenta \textsc{Yamamura}\altaffilmark{14},
        Lewis M. \textsc{Cook}\altaffilmark{15},
        Richard \textsc{Sabo}\altaffilmark{16},
        Ian \textsc{Miller}\altaffilmark{17},
        William \textsc{Goff}\altaffilmark{18},
        Seiichiro \textsc{Kiyota}\altaffilmark{19},
        Sergey Yu. \textsc{Shugarov}\altaffilmark{11,20},
        Polina \textsc{Golysheva}\altaffilmark{11},
        Olga \textsc{Vozyakova}\altaffilmark{11}, 
        Stephen M. \textsc{Brincat}\altaffilmark{21}, 
        Hiroshi \textsc{Itoh}\altaffilmark{22},
        Tam\'as \textsc{Tordai}\altaffilmark{23},
        Colin \textsc{Littlefield}\altaffilmark{24},
        Roger D. \textsc{Pickard}\altaffilmark{25,26},
        Kenji \textsc{Tanabe}\altaffilmark{27},
        Kenzo \textsc{Kinugasa}\altaffilmark{28,29}, 
        Satoshi \textsc{Honda}\altaffilmark{28,30}, 
        Hikaru \textsc{Taguchi}\altaffilmark{28}, 
        Osamu \textsc{Hashimoto}\altaffilmark{28}, and 
        Daisaku \textsc{Nogami}\altaffilmark{1}
        }
\email{mkimura@kusastro.kyoto-u.ac.jp}

\altaffiltext{1}{Department of Astronomy, Graduate School 
of Science, Kyoto University, Oiwakecho, Kitashirakawa, 
Sakyo-ku, Kyoto 606-8502}

\altaffiltext{2}{Okayama Astrophysical Observatory, National 
Astronomical Observatory of Japan, 3037-5 Honjo, Kamogata, 
Asakuchi, Okayama, 719-0232}

\altaffiltext{3}{Okayama Observatory, Graduate School of 
Science, Kyoto University, 3037-5 Honjo, Kamogata, Asakuchi, 
Okayama 719-0232}

\altaffiltext{4}{Institute of Astronomy and Astrophysics, 
Academia Sinica, 11F of Astronomy-Mathematics Building, 
AS/NTU No. 1, Section 4, Roosevelt Road, Taipei 10617, Taiwan}

\altaffiltext{5}{Subaru Telescope, National Astronomical 
Observatory of Japan, 650 North A'ohoku Place, Hilo, HI 96720, USA}

\altaffiltext{6}{Bronberg Observatory, Center for Backyard Astrophysics Pretoria, PO Box 11426, 
Tiegerpoort 0056, South Africa}

\altaffiltext{7}{Kleinkaroo Observatory, Center for Backyard Astrophysics Kleinkaroo, Sint Helena 1B, PO Box 281, Calitzdorp 6660, South Africa}

\altaffiltext{8}{Variable Star Observers League in Japan (VSOLJ), 124 Teradani, Isato-cho, Kumano, Mie 519-4673}

\altaffiltext{9}{American Association of Variable Star Observers, 
49 Bay State Rd., Cambridge, MA 02138, USA}

\altaffiltext{10}{Crimean Astrophysical Observatory, 
298409, Nauchny, Republic of Crimea}

\altaffiltext{11}{Sternberg Astronomical Institute, Lomonosov 
Moscow State University, Universitetsky Ave., 13, Moscow 119992, 
Russia}

\altaffiltext{12}{Physics Department, Rochester Institute of 
Technology, Rochester, New York 14623, USA}

\altaffiltext{13}{Instituto de Astronom\'{i}a UNAM, Apartado Postal 877, 22800 Ensenada B.C., M\'{e}xico}

\altaffiltext{14}{Osaka Kyoiku University, 4-698-1 Asahigaoka, 
Kashiwara, Osaka 582-8582}

\altaffiltext{15}{Center for Backyard Astrophysics Concord, 1730 
Helix Ct. Concord, California 94518, USA}

\altaffiltext{16}{2336 Trailcrest Dr., Bozeman, Montana 59718, USA}

\altaffiltext{17}{Furzehill House, Ilston, Swansea, SA2 7LE, UK}

\altaffiltext{18}{American Association of Variable Star Observers 
(AAVSO), 13508 Monitor Lane, Sutter Creek, California 95685, USA}

\altaffiltext{19}{Variable Star Observers League in Japan (VSOLJ), 
7-1 Kitahatsutomi, Kamagaya, Chiba 273-0126}

\altaffiltext{20}{Astronomical Institute of the Slovak Academy of 
Sciences, 05960 Tatranska Lomnica, Slovakia}

\altaffiltext{21}{Flarestar Observatory, San Gwann SGN 3160, Malta}

\altaffiltext{22}{Variable Star Observers League in Japan (VSOLJ), 
1001-105 Nishiterakata, Hachioji, Tokyo 192-0153}

\altaffiltext{23}{Polaris Observatory, Hungarian Astronomical 
Association, Laborc utca 2/c, 1037 Budapest, Hungary}

\altaffiltext{24}{Astronomy Department, Wesleyan University, Middletown, 
Connecticut 06459, USA}

\altaffiltext{25}{The British Astronomical Association, 
Variable Star Section (BAA VSS), Burlington House, Piccadilly, 
London, W1J 0DU, UK}

\altaffiltext{26}{3 The Birches, Shobdon, Leominster, Herefordshire, 
HR6 9NG, UK}

\altaffiltext{27}{Department of Biosphere-Geosphere System Science, Faculty of Informatics, Okayama University of Science, 1-1 Ridai-cho, Okayama, Okayama 700-0005}

\altaffiltext{28}{Gunma Astronomical Observatory, 6860-86 Nakayama, 
Takayama, Agatsuma, Gunma 377-0702}

\altaffiltext{29}{Nobeyama Radio Observatory, National Astronomical Observatory of Japan, National Institutes of Natural Sciences, 462-2 Nobeyama, Minamimaki, Minamisaku, Nagano 384-1305}

\altaffiltext{30}{Nishi-Harima Astronomical Observatory, Center for Astronomy, University of Hyogo, 407-2 Nishigaichi, Sayo-cho, Sayo, Hyogo 679-5313}

\title{On the Nature of Long-Period Dwarf Novae with Rare and Low-Amplitude Outbursts}

\Received{} \Accepted{}


\SetRunningHead{Kimura et al.}{On the Nature of Long-Period DNe with Rare and Low-Amplitude Outbursts}

\maketitle

\begin{abstract}
There are several peculiar long-period dwarf-nova like objects, 
which show rare, low-amplitude outbursts with highly ionized 
emission lines.  1SWASP J162117$+$441254, BD Pav, and V364 Lib 
belong to this kind of objects.  Some researchers even doubt 
whether 1SWASP J1621 and V364 Lib have the same nature as normal 
dwarf novae.  
We studied the peculiar outbursts in these three objects via 
our optical photometry and spectroscopy, and 
performed numerical modeling of their orbital variations to 
investigate their properties.  
We found that their outbursts lasted for a long interval 
(a few tens of days), and that slow rises in brightness 
were commonly observed during the early stage of their outbursts.  
Our analyses and numerical modeling suggest that 1SWASP 
J1621 has a very high inclination, close to 90 deg, plus 
a faint hot spot.  Although BD Pav seems to have a slightly 
lower inclination ($\sim$75 deg), the other properties are 
similar to those in 1SWASP J1621.  On the other hand, V364 Lib 
appears to have a massive white dwarf, a hot companion star, and 
a low inclination ($\sim$35 deg).  
In addition, these three objects possibly have low transfer 
rate and/or large disks originating from the long orbital 
periods.  
We find that these properties of the three objects can 
explain their infrequent and low-amplitude outbursts 
within the context of the disk instability model in normal 
dwarf novae without strong magnetic field.  
In addition, we suggest that the highly-ionized emission 
lines in outburst are observed due to a high inclination 
and/or a massive white dwarf.  
More instances of this class of object may be unrecognized, 
since their unremarkable outbursts can be easily overlooked.  
\end{abstract}

\section{Introduction}

   Statistical studies on dwarf novae show many dwarf novae 
having long orbital periods (more than 3 hours) go through 
outbursts with typical amplitudes of 2--5 mag, and that the 
intervals between their outbursts are usually less than 
1 year (see \cite{war95book} for a review).  
It is widely accepted that dwarf novae showing low outburst 
amplitudes undergo outbursts more frequently 
\citep{war87CVabsmag}.  
A very low-amplitude and rare outburst was, however, discovered 
for the first time in a dwarf-nova like object, 
ASAS 150946$-$2147.7 = V364 Lib, \citep{poj09asas1509}.  
This object has a long orbital period of 16.86 hours, and 
enters its outbursts with amplitudes of $\sim$1 mag every few years 
(\cite{wil11j1714asas1509}; vsnet-alert 14271; vsnet-alert 20877).  
In addition, highly ionized emission lines (He II 4686 and 
C III/N III) were detected in the 2009 outburst.  
Although this led \citet{kin09asa1509cbet} to propose that this 
system is a black-hole binary, \citet{uem09asas1509atel} 
reported that the X-ray luminosity during the outburst was 
comparable to the quiescent X-ray luminosity of black-hole binaries.  
Thus this object contradicts the general rule of dwarf novae, 
and the origin of the peculiar outburst and highly ionized 
emission lines remains a mystery.  

   On June 3.45, 2016 (UT), a small-amplitude and very rare 
outburst was discovered in another object by Catalina Real-Time 
Transient Survey (CRTS) \citep{dra16j162117,dra14CRTSCVs}.  
Its name is 1SWASP J162117$+$441254 (hereafter 1SWASP J1621), 
and this object also has a long orbital period of 4.99 hours 
\citep{loh13swasp}.   
The outburst amplitude was $\sim$2 mag and only one outburst 
was detected over ten years before the 2016 outburst 
in this system \citep{kju17j1621}.  The strong He II 4686 
emission line observed in the outburst was similar to that in 
the 2009 outburst of V364 Lib \citep{sca16j162117}.  
Moreover, this object had been previously identified as 
a W UMa-type system on the basis of SDSS colors and 
double-waved orbital variations in quiescence 
\citep{pal13periodicLC,loh13eclipse,dra14catalog}; 
hence, the 2016 outburst was expected to be the onset of 
a merger of two main-sequence (MS) stars, which was reminiscent 
of the 2008 eruption in V1309 Sco.  
\citet{sca16j162117}, however, confirmed by their spectroscopic 
observations that this outburst occurred in an accreting compact 
binary rather than a MS$+$MS merger.  
In addition, this system showed much deeper eclipses in outburst 
than in quiescence 
(\cite{zej16j162117,kju17j1621,zol17j1621}; Sec.~3.1 in this paper).  
This phenomenon made us notice the similarity between this object 
and the almost forgotten dwarf nova BD Pav, which has an orbital 
period of 4.3 hours \citep{bar83bdpav}.  BD Pav has also shown 
relatively low-amplitude and rare outbursts, and the He II 4686 
emission line was observed in its 1985 outburst.  
The emission line was not clearly double-peaked, in spite of 
the deep eclipses in outburst suggesting its high 
inclination \citep{bar87bdpav}.  

   Recently, the two objects GY Hya and V1129 Cen, which 
showed outbursts, spectra, and orbital variability similar to 
those in 1SWASP J1621, BD Pav and V364 Lib, have been studied 
\citep{bru17gyhya,bru17v1129cen}.  
In addition, HS 0218$+$3229 (hereafter HS 0218) shows rare 
outbursts and has a long orbital period, although the outburst 
amplitude is not so small, $\sim$4 mag in the $V$ band 
\citep{gol12hs0218,gol13hs0218,kat15hs0218}, and 
it was pointed out that this object is similar to 1SWASP J1621 
\citep{kat17j1621}.  
Thus this kind of object recently has started to attract 
attention, but there is not yet any coherent explanation to 
produce its peculiar outburst and spectral behavior.  
Moreover, some of these objects were not clearly identified 
with normal dwarf novae, and it is even doubted whether 
the disk instability model, which is believed to be 
the most plausible model of outbursts in non-magnetic 
dwarf novae (\cite{osa96review} for a review), can explain 
the outbursts in this kind of object.  
For example, Dr.~Breedt claims that 1SWASP J1621 has a hot 
companion star \citep{waa17j1621}.  Although this assumption 
would be true in V1129 Cen according to \citet{bru17v1129cen}, 
it is not consistent with the spectroscopic observations of 
1SWASP J1621 and BD Pav \citep{tho16j162117,sio08CVHST,fri90CVsodium}.  
In addition, \citet{qia17j1621} even argued that the eruption 
in 1SWASP J1621 was not a dwarf-nova outburst.  They considered 
the accretion disk is barely formed due to the strong magnetic 
activity of the companion star in this system, and that a sudden 
increase of mass transfer from the companion causes the eruption.  
We therefore aimed to investigate the nature of 1SWASP J1621, 
BD Pav, and V364 Lib, and to examine by analyzing our photometric 
and spectroscopic observational data and modeling their orbital 
variations, whether some special feature which is not presented 
in normal dwarf novae is necessary to explain their peculiar 
outbursts.  

   The plan of this paper is as follows: the data reduction and 
techniques of analyses are described in Sec.~2.
We present our optical photometry of 1SWASP J1621, BD Pav, and 
V364 Lib in Sec.~3, and optical spectroscopy of V364 Lib 
in Sec.~4.  In addition, we perform numerical modeling to 
explain the orbital variations in Sec.~5.  
In Sec.~6, we discuss our results, and in Sec.~7, our conclusions 
are summarized.

\section{Observation and Analysis}

\subsection{Photometry}

   Time-resolved CCD photometry was carried out by the VSNET 
collaboration team.  The telescopes and sites are summarized 
in Table E1. Tables E2, E3, E4 show the logs of photometric 
observations of 1SWASP J1621, BD Pav, and V364 Lib, respectively.  
We also used the data downloaded from the AAVSO 
archive\footnote{$<$http://www.aavso.org/data/download/$>$}, 
the ASAS-3 data archive \citep{poj04asas} and 
the ASAS-SN data archive \citep{dav15ASASSNCVAAS}.
All of the observation times were converted to barycentric 
Julian date (BJD).
Before making the analyses, we applied zero-point corrections 
to each observer by adding constants as for 1SWASP J1621 and 
V364 Lib.  We did not do that as for BD Pav,  since all of 
the observations were performed by one observer.  
The magnitude scales of each site were adjusted to that of 
the Crimean Observatory system (CRI in Table E2), where 
TYC 3068-00855-1 (RA: 16h21m05.42s, Dec:+\timeform{44D14'32.8''}, 
$V$ = 12.2) was used as the comparison star for 1SWASP J1621.  
The constancy of the comparison star was checked by nearby 
stars in the same images.  The magnitude of the comparison 
star was measured by the AAVSO Photometric All-Sky Survey 
(APASS: \cite{APASS}) from the AAVSO Variable Star 
Database.\footnote{$<$http://www.aavso.org/vsp$>$}
As for V364 Lib, we adjusted the magnitude of our 
observational data to that of the ASAS data, assuming the 
magnitude in a clear filter is almost the same as that in 
the $V$ band.  
The exposure times of the observations by the VSNET team 
were 30--300 s as for 1SWASP J1621 and V364 Lib, and 30 s 
as for BD Pav.

\subsection{Period Analyses}

   We used the phase dispersion minimization (PDM) method 
\citep{PDM} for period analyses.  We subtracted the global 
trend of the light curve by locally weighted polynomial 
regression (LOWESS: \cite{LOWESS}) before performing the PDM 
analysis.  The 1$\sigma$ error of the best estimated period 
by the PDM analysis was determined by the methods in 
\citet{fer89error} and \citet{Pdot2}.  
   A variety of bootstraps was used for estimating the 
robustness of the result of PDM.  We analyzed about 100 samples 
which randomly contain 50\% of observations, and performed 
a PDM analysis for these samples.

\subsection{Spectroscopy}

   Our spectroscopic observations of V364 Lib were 
carried out by using the Gunma Astronomical Observatory's 
1.5-m telescope equipped with Gunma LOW resolution Spectrograph 
and imager (GLOWS) in the 2009 outburst and using the High 
Dispersion Spectrograph (HDS: \cite{nog02subaru}) attached to 
the 8.2-m Subaru telescope of National Astronomical Observatory 
of Japan (NAOJ) in quiescence just after the outburst.  
The spectral coverage of the two spectrographs was about 
4000--8000 and 3000--10000 \AA, respectively, 
and the spectroscopic resolutions ($R=\lambda/\Delta \lambda$) 
are 400--500 for GLOWS and 90,000 for HDS.  
Data reduction was conducted using IRAF\footnote{IRAF is 
distributed by the National Optical Astronomy Observatories, 
which is operated by the Association of Universities for Research 
in Astronomy, Inc., under cooperate agreement with the National 
Science Foundation.} in the standard manner (bias subtraction, 
flat fielding, aperture determination, scattered light subtraction, 
spectral extraction, wavelength calibration, normalization by the 
continuum, and heliocentric radial-velocity correction).  
We measured the radial velocities of the object by fitting 
gaussians with the task SPLOT in IRAF.  
The observation dates, exposure times, and observatories are 
summarized in Table E5.  
\textcolor{black}{The estimated radial velocities are given 
in Tables E6 and E7}.

\section{Photometric Light Curves}

\subsection{1SWASP J1621}

\begin{figure}[htb]
\begin{center}
\FigureFile(80mm, 50mm){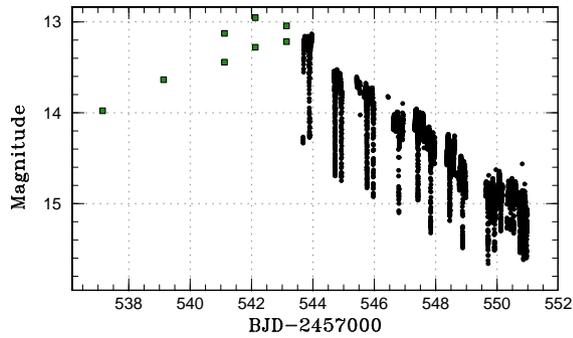}
\end{center}
\caption{Overall light curve of the 2016 outburst in 1SWASP J1621 (BJD 2457537--2457551).  The filled rectangles represent the snap-shot observations by Hiroyuki Maehara \citep{mae16j162117}.  }
\label{j162117-overall}
\end{figure}

\begin{figure}[htb]
\begin{center}
\FigureFile(60mm, 50mm){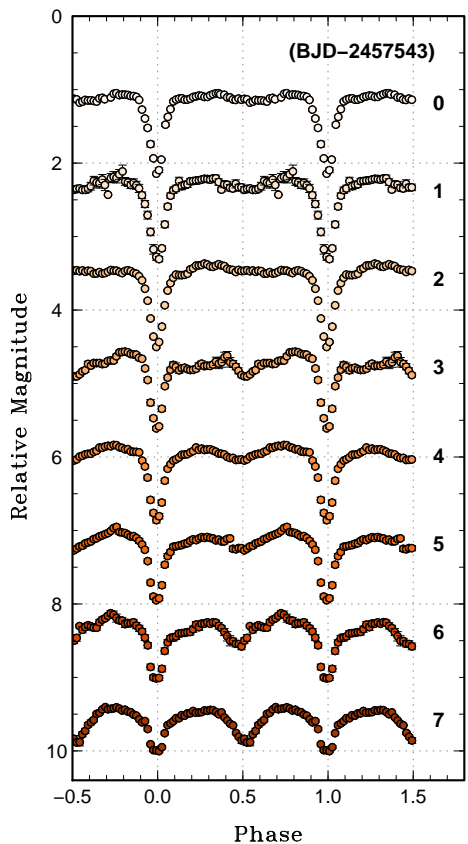}
\vspace{1cm}
\end{center}
\caption{Nightly eclipsing variations in magnitudes with clear filter in 1SWASP J1621 during the 2016 outburst (BJD 2457543--2457550).  The numbers at the right end represent the days from the beginning of the outburst.  }
\label{j162117-daily}
\end{figure}

\begin{figure}[htb]
\begin{minipage}{1.0\hsize}
\begin{center}
\FigureFile(70mm, 100mm){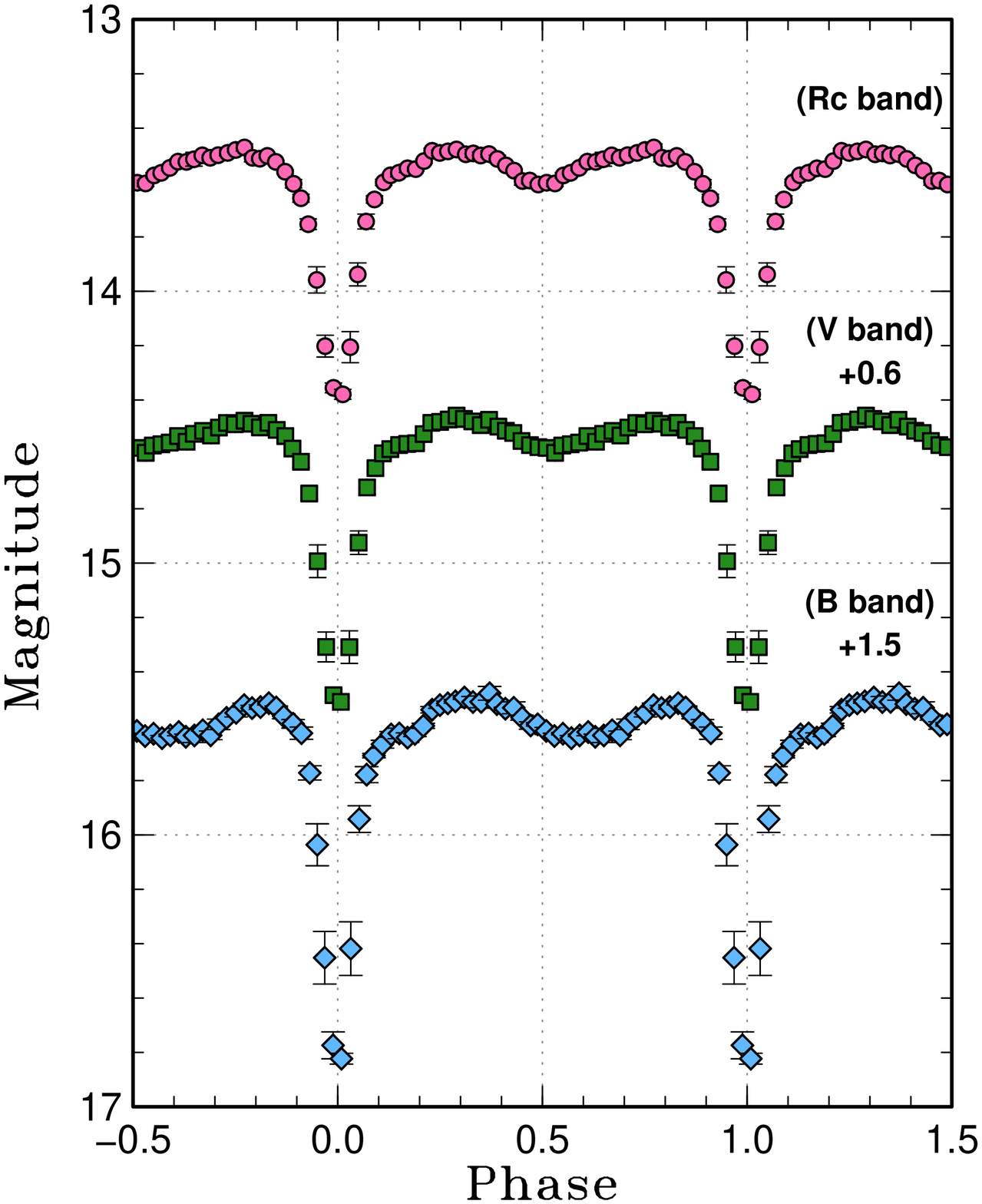}
\vspace{-5mm}
\end{center}
\end{minipage}
\\
\begin{minipage}{1.0\hsize}
\begin{center}
\FigureFile(70mm, 50mm){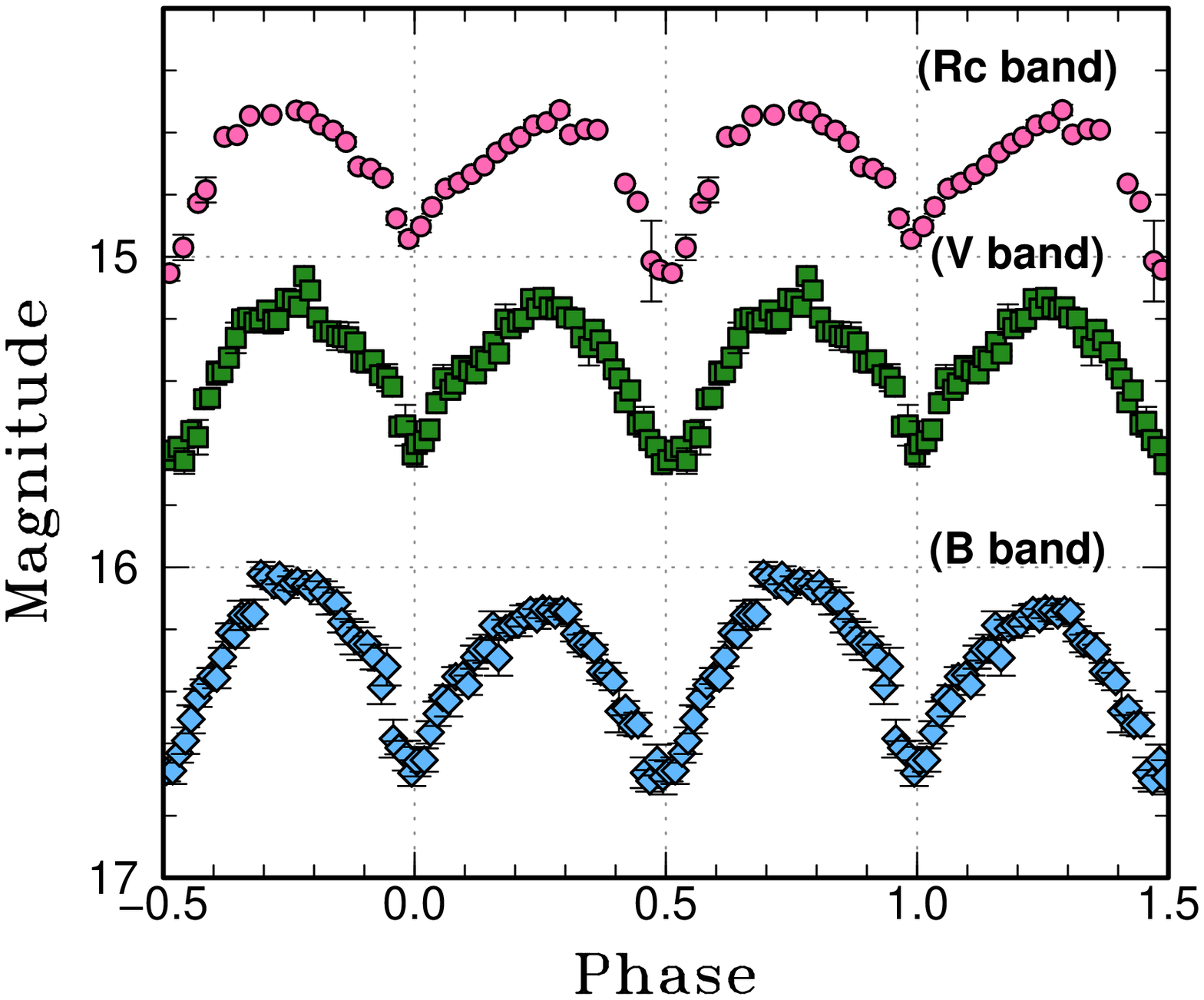}
\vspace{5mm}
\end{center}
\end{minipage}
\caption{Phase-averaged light curves of orbital variations in the outburst state during BJD 2457542.8--2457546.8 (the upper panel) and in quiescence (the lower panel) during BJD 2457552--2457559 in 1SWASP J1621.  Diamonds, rectangles, and circles represent the $B$, $V$ and $R_{\rm C}$-bands phase profiles, respectively.  In the upper panel, the $V$-band and $B$-band magnitudes are offset by 0.6 and 1.5, respectively, for visibility.  The folding period is 0.207852 d, which was reported by \citet{dra14CRTSCVs}.  The epochs in the outburst maximum and in quiescence are BJD 2457546.59 and BJD 2457549.7079, respectively.  }
\label{phase}
\end{figure}

   In the 2016 outburst, the amplitude of its outburst was 
relatively small, $\sim$2 mag.  The duration of the outburst 
was probably about two weeks, and the brightness rose slowly 
in the early stage of the event.  
The overall optical light curve of the 2016 outburst in 
1SWASP J1621 with a clear filter is displayed in figure 
\ref{j162117-overall}.  
This system showed deep eclipsing variations on its light curve.  
As this system was fading, the primary minima became shallower 
and the secondary minima became deeper.  
We display in figure \ref{j162117-daily} the nightly averaged 
phase profiles in a clear filter during the fading stage 
in the outburst.  
This system showed W UMa-type orbital variations with amplitudes 
of $\sim$0.6 mag in quiescence and deep primary minima in outburst.  
The phase-averaged profiles around the outburst maximum and 
in quiescence during BJD 2457552--2457559 in the $R_{\rm C}$, 
$V$, and $B$ bands are exhibited in figure \ref{phase}.  
The profile around the outburst maximum was derived from 
the multi-color observational data obtained by Mic during 
BJD 2457542.8--2457546.8 (see the annotations in Table E2), 
which include 3 cycles of the eclipses.  
When folding the outburst light curve, we subtracted 
the long-term trend by using locally weighted polynomial 
regression (LOWESS: \cite{LOWESS}) after removing the part 
of the phase representing the deep eclipse.  
The derived outburst amplitude is $\sim$1 mag in the $R_{\rm C}$ 
band from the phase-averaged light curves.  
During the outburst, we observed colors of $V-R \sim 0.4$ mag 
and $B-V \sim 0.1$ mag outside eclipses; in quiescence, those 
colors were $V-R \sim 0.6$ mag and $B-V \sim 1.0$ mag.  
In the $B$ band, the flux of the maxima around phase 0.25 
was a little lower than that of the maxima around phase 0.75.  

\subsection{BD Pav}

\begin{figure}[htb]
\begin{center}
\FigureFile(80mm, 50mm){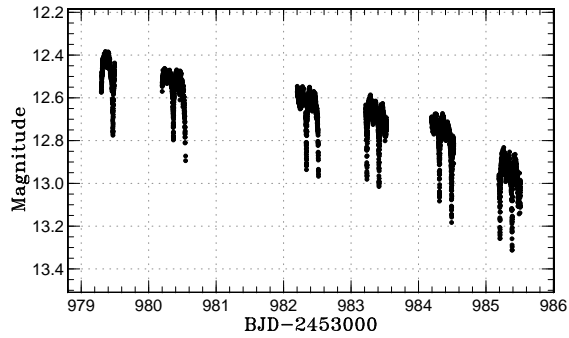}
\end{center}
\caption{Overall light curve of the 2006 outburst in BD Pav with a clear filter (BJD 2453979--2453986).  }
\label{bdpav-overall}
\end{figure}

\begin{figure}[htb]
\begin{center}
\FigureFile(70mm, 50mm){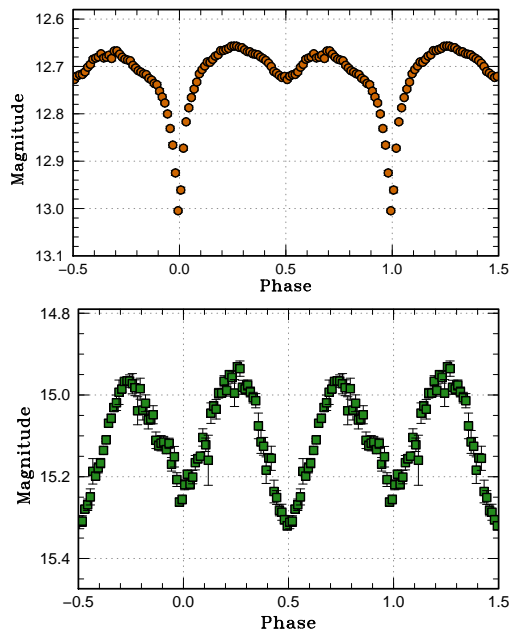}
\end{center}
\caption{Phase-averaged profiles of orbital variations of BD Pav in the 2006 outburst during BJD 2453979--2453986 (the upper panel) and in quiescence during BJD 2456454--2456466 (the lower panel).  The folding period is 0.17930, which is reported in \citet{bar83bdpav}.  The epochs are BJD 2453982.339 in the outburst and BJD 2456454.725 in quiescence.  }
\label{bdpav-phase}
\end{figure}

\begin{figure}[htb]
\begin{center}
\FigureFile(80mm, 50mm){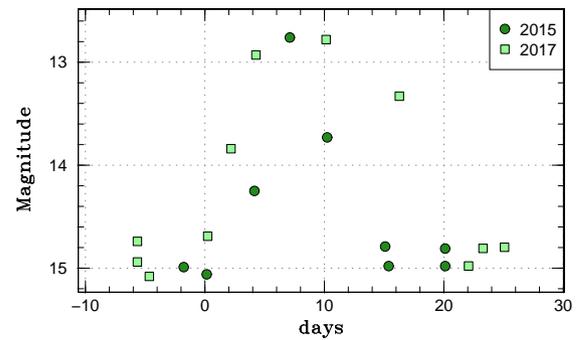}
\end{center}
\caption{Monitoring BD Pav in the $V$ band by ASAS during the 2015 and 2017 outbursts.  The horizontal axis represents time in days from each outburst.  The circles and squares represent the 2015 and 2017 outbursts, respectively.}
\label{bdpav-2017}
\end{figure}

   We performed the optical photometry of the 2006 outburst 
of BD Pav in the past.  The outburst amplitude was 
$\sim$2.5 mag.  The overall light curve of this outburst 
is shown in figure \ref{bdpav-overall}.  
In addition, we confirmed the deep primary minima in this 
outburst similar to those in the 2016 outburst of 1SWASP J1621.  
In figure \ref{bdpav-phase}, we show the phase-averaged 
profiles both in the 2006 outburst and in quiescence.  
In folding the outburst light curve, we subtracted 
the long-term trend using the same method described for 
1SWASP J1621.
We can see that this object showed the much deeper primary 
minima in the outburst than in quiescence and that 
double-waved orbital modulations with amplitudes of 
$\sim$0.3 mag were observed in quiescence.  
\textcolor{black}{
The main characteristics of these eclipsing profiles are 
consistent with the results in Figures 2 and 3 
in \citet{bar87bdpav}.  }
Although the rising phase in the 2006 outburst was not observed 
in our photometry, many other outbursts in BD Pav showed slow 
rises.  We show measurements in the $V$ band by ASAS during 
the recent 2015 and 2017 outbursts in figure \ref{bdpav-2017} 
as examples.  

\subsection{V364 Lib}

\begin{figure}[htb]
\begin{center}
\FigureFile(80mm, 50mm){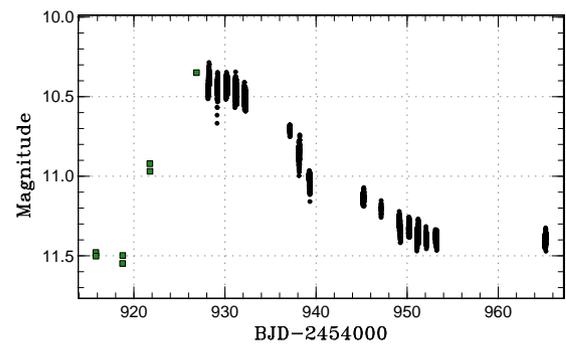}
\end{center}
\caption{Overall light curve of the 2009 outburst in V364 Lib with a clear filter (BJD 2454928--2454954).  The squares represent observations in the $V$ band by ASAS.  }
\label{asas1509-overall}
\end{figure}

\begin{figure}[htb]
\begin{center}
\FigureFile(70mm, 50mm){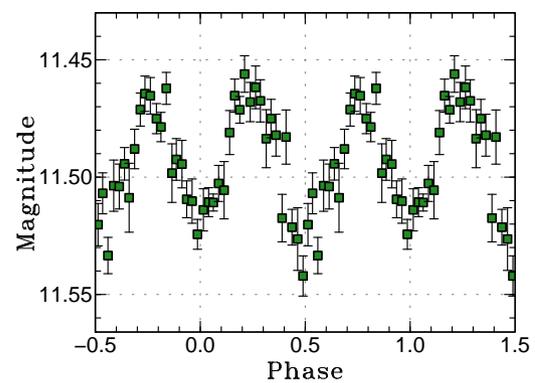}
\end{center}
\caption{Phase-averaged profile of orbital variations in quiescence in V364 Lib.  The folding orbital period estimated by the PDM method is 0.7024293(1053) d.  The epoch is BJD 2453880.244654.  }
\label{phase-asas1509}
\end{figure}

\begin{figure}[htb]
\begin{center}
\FigureFile(80mm, 50mm){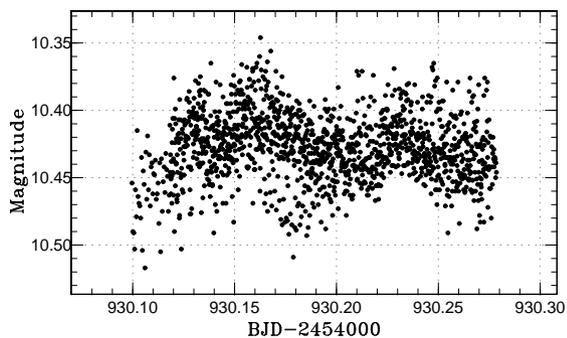}
\end{center}
\caption{An example of the variations on timescales shorter than the orbital period in the 2009 outburst in V364 Lib.  }
\label{shortvari}
\end{figure}

   We carried out the optical photometry of the 2009 outburst 
of V364 Lib in the past.  The duration was about 35 days and the 
outburst amplitude was $\sim$1 mag.  A slow rise during the early 
stage of the outburst was observed, as in the outbursts of 1SWASP 
J1621 and BD Pav.  Its overall light curve is displayed in figure 
\ref{asas1509-overall}.  
Small-amplitude ($\sim$0.07 mag) and double-waved orbital variations 
were detected in quiescence, as displayed in figure \ref{phase-asas1509}.  
In outburst, these orbital variations were expected to be observed 
with an amplitude of $\sim$0.025 mag; however they were not detected 
due to other small-amplitude variations on time scales of $\sim$0.1 d, 
which are shorter than the orbital period (and below the limits 
of our precision).  
A part of the variations in outburst is displayed in figure 
\ref{shortvari}.  

\section{Spectroscopy in V364 Lib}

\begin{figure}[htb]
\begin{center}
\FigureFile(80mm, 50mm){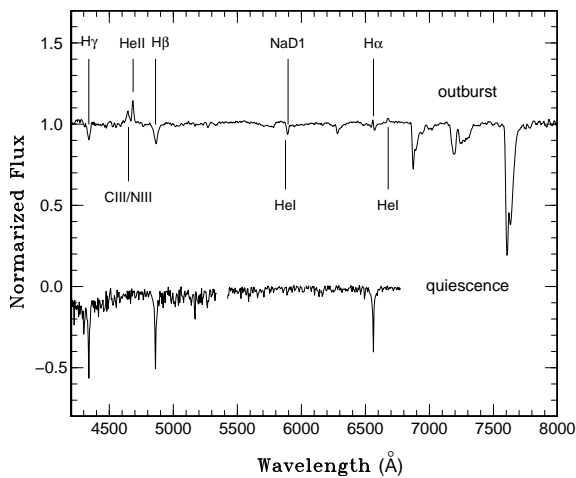}
\end{center}
\caption{Spectrum of V364 Lib on April 7th, 2009 in the outburst state.  For reference, an example of the spectra in quiescence on May 5th, 2009 is also displayed.  For visibility, an offset of $-$1.0 is added to the quiescent spectrum.}
\label{asas1509-spec}
\end{figure}

\begin{figure}[htb]
\begin{center}
\FigureFile(100mm, 60mm){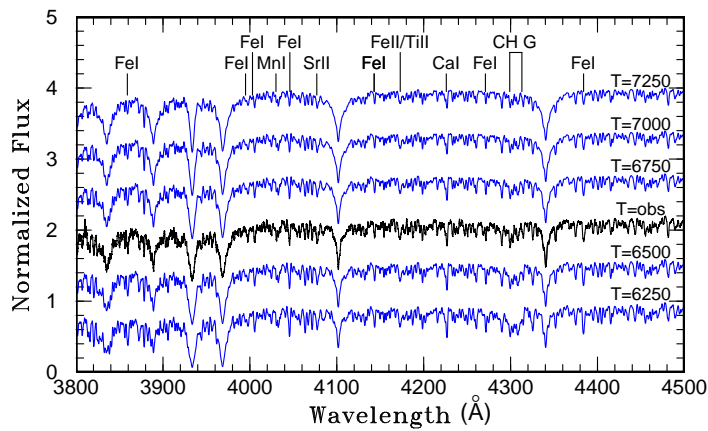}
\end{center}
\caption{Spectrum of V364 Lib in quiescence.  The black line represents observations on May 5th, 2009.  The blue lines represent the broadened synthetic spectra in \citet{cas04atlas9}.  }
\label{temp2}
\end{figure}

   We detected absorption lines in the Balmer series and 
strong emission lines of He II 4686 and C III/N III during 
the early stage of the 2009 outburst in V364 Lib.  
An example of optical spectra in the early stage of the 
outburst is given in figure \ref{asas1509-spec} with the 
spectrum in quiescence for comparison.  
In quiescence, the contribution of the companion star 
was dominant in the optical spectra.  
The companion star was determined to be an F-type star with 
surface temperature between 6,500 K and 6,750 K 
by comparing the observational spectra with the synthetic 
spectra which are computed with a synthetic stellar atmosphere 
interpolated from the \citet{cas04atlas9} 
grid\footnote{https://www.oact.inaf.it/castelli/castelli/grids.html}.  
An example of optical spectra in quiescence is given in figure 
\ref{temp2} with some synthetic spectra.

\begin{figure}[htb]
\begin{center}
\FigureFile(80mm, 50mm){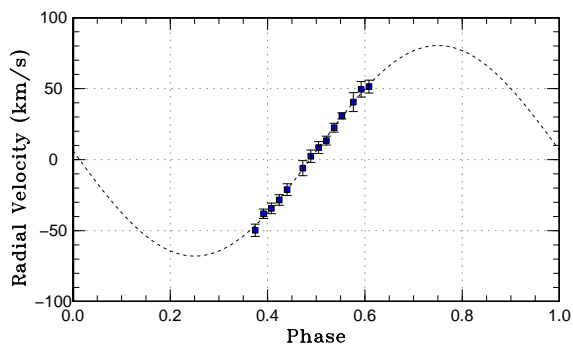}
\end{center}
\caption{Radial velocity extracted from the average value of H$\beta$ and Mg absorption lines in quiescence under the assumption that the orbital period is 0.7024293 d.  The rectangles represent the observations.  The dashed line is the best fitted sine curve with the semi-amplitude of 74.1$\pm$1.1 km/s and the systemic velocity of 6.2$\pm$5.4 km/s.  }
\label{rvk2}
\end{figure}

\begin{figure}[htb]
\begin{center}
\FigureFile(80mm, 50mm){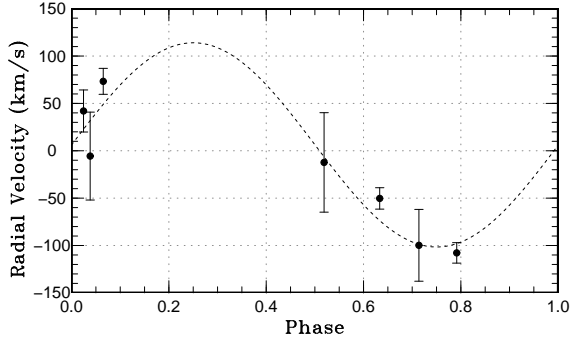}
\end{center}
\caption{Radial velocity extracted from He II 4686 emission line in the 2009 outburst under the assumption that the orbital period is 0.7024293 d.  The points represent the observations.  The dashed line is the best fitted sine curve with the semi-amplitude of 107.8$\pm$12.6 km/s and the systemic velocity of 6.2 km/s.  }
\label{rvk1}
\end{figure}

   We deduced that the central compact object in this system 
is a white dwarf as follows.  
The estimated value of the radial velocity of the companion 
star (K2), using the average value of H$\beta$ and Mg 
absorption lines in quiescence, was 74.1$\pm$1.3 km/s (see 
figure \ref{rvk2}). 
Although the error was somewhat large, the derived value of 
the radial velocity of the central compact object (K1) from 
the He II emission line in the outburst state was 
107.8$\pm$12.6 km/s (see figure \ref{rvk1}).  
We then assumed that the systemic velocity is the same as 
that in quiescence, and that the orbital period is 0.7024293 d.  
Thus the mass ratio \textcolor{black}{
($q \equiv M_2/M_1$)\footnote{\textcolor{black}{Here, $M_1$ and $M_2$ 
represent the mass of the primary star and that of the 
secondary star, respectively.}}} 
was estimated to be $1.5 \pm 0.2$.  
The surface temperature of the companion suggests that its mass 
is close to 1.4 $M_{\solar}$ \citep{all73quantities}, and hence, 
the mass of the compact object was estimated to be $1.0 \pm 0.2 
M_{\solar}$ from the mass ratio.  
This value and the faint X-ray luminosity during outburst 
\citep{uem09asas1509atel} indicate that 
the compact object in this system should be a white dwarf.  
In addition, the inclination angle was estimated to be 
$\sim$35 deg from the equation $\sin^{3}i = P {K_{2}}^{3} 
(1+q)^{2}/(2 \pi G M_{1})$, using the estimated values of K2, 
mass ratio, and mass of the primary, which is consistent with 
a $v\sin{i}$ value of 85 km/s estimated from the absorption 
lines in quiescence.  

\section{Numerical Modeling of Orbital Variations}

\subsection{Premises}

As described in the introduction, the observational 
W UMa-type orbital variations in quiescence are sometimes 
ascribed to their hot companions in 1SWASP J1621 and 
BD Pav \citep{waa17j1621,bar83bdpav}, but there is evidence 
that 1SWASP J1621 and BD Pav have cool companion stars 
\citep{tho16j162117,sio08CVHST,fri90CVsodium}.  
In addition, our spectral analyses in Sec.~4 suggest 
that V364 Lib has a hot companion star and a low 
inclination.  
We thus wondered if the double-waved modulations 
in quiescence and the outburst amplitudes are explained 
without hot companions as for 1SWASP J1621 and BD Pav, 
and with the nature suggested by our results in Sec.~4 
in V364 Lib, via the modeling of the orbital phase 
profiles.  
Since we have the $R_{\rm C}$, $V$, and $B$-band phase 
profiles in outburst (see figure \ref{phase}) in 1SWASP J1621, 
we considered reproducing their characteristics 
in this object.  

Before presenting our modeling, we summarize the information 
derived from our observations.  
According to \citet{EclipseMapping}, the width of the eclipse of 
a white dwarf is determined by the mass ratio and inclination 
of its system.  
The width is approximately estimated as the width of eclipse 
at half depth if the disk is axisymmetric.  
Regarding the inflection phases in outburst in the 
$R_{\rm C}$-band profile for 1SWASP J1621 and in the 
clear-filter profile for BD Pav as the ends of eclipses, 
the rough estimates of the width of eclipse of the white dwarf 
are $\sim$0.10 in 1SWASP J1621 and $\sim$0.06 in BD Pav, 
respectively.  
Substituting the half values of them and the values of mass 
ratios to equation (3) in \citet{hor82v363aur}, 
which represents the relation between the inclination, mass 
ratio, and width of eclipse of a white dwarf, we obtain 
inclinations of $\sim$87-deg in 1SWASP J1621 and $\sim$75-deg 
in BD Pav, respectively.  These values are 
consistent with those extracted by 
\citet{kju17j1621,zol17j1621,fri90CVsodium,sio08CVHST}.  
In addition, the almost symmetric phase-averaged profiles 
around the primary minima in these two objects seem to imply 
a dim hot spot (see figures \ref{phase} and \ref{bdpav-phase}).

\subsection{Method}

\subsubsection{Code}

   We used the numerical code which is described in detail by 
\citet{hac01RN}, \citet{hac03SSS}, and \citet{hac03j0513}.  
In this code, the companion star fills its Roche lobe 
and a circular orbit is assumed.  The surfaces of the white 
dwarf, companion star, and accretion disk are divided into 
patches, and each patch emits photons by black body 
radiation at the local temperature.  
The patches of the disk and companion are irradiated by 
the front-side patches of the white dwarf if there is no patch 
between them.  The non-irradiated surface temperature is 
determined by viscous heating of the standard disk model 
\citep{sha73BHbinary}.  The total luminosity of the binary 
system is calculated by summing up the luminosity from all 
visible patches.  The surface patch elements are 32$\times$64 
($\theta \times \phi$) for the companion, 32$\times$64 ($\theta 
\times \phi \times$ up and down side) for the accretion disk, 
and 16$\times$32 for the white dwarf. 

\subsubsection{Parameters}

   Using the code described in Sec.~5.2.1, we computed 
eclipsing light variations with various test 
parameters: the inclination angle, temperature distribution, 
height and radius of the accretion disk.  
Inclinations were varied over ranges of 84--90 deg, 72--78 
deg, and 30--40 deg by steps of 1 deg for 1SWASP J1621 and 
BD Pav, V364 Lib, respectively, by considering the estimation 
in Sec.~4 and Sec.~5.1.   
Tested disk radii were 0.5, 0.6, 0.7, 0.8, 0.9 $R_{\rm L1}$.  
Here, $R_{\rm L1}$ is defined as $R_{\rm L1} = a / (1.0015 + 
q^{0.4056})$, where $a$ is the binary separation 
(equation (2.4c) in \cite{war95book}).  

   We gave the temperature distribution of the accretion disk 
in outburst, postulating that the disk is steady at the 
outburst maximum \citep{woo86zcha} as follows: 
\begin{eqnarray}
T = T_{*} \left(\frac{r}{r_{\rm in}} \right)^{-3/4}\left[1-\left(\frac{r_{\rm in}}{r} \right)^{1/2} \right]^{1/4}, 
T_{*} = \left(\frac{3 G M \dot{M}}{8 \pi \sigma r_{\rm in}^{3}} \right)^{1/4}. 
\label{temp-steady}
\end{eqnarray}
Here, $r$, $r_{\rm in}$, $G$, $M$, $\dot{M}$ and $\sigma$ 
represent the distance from the central object to a given ring 
of the disk, inner disk 
radius, constant of gravitation, mass of the primary star, 
mass accretion rate at a given radius, and Stefan-Boltzmann 
constant, respectively \citep{sha73BHbinary}, 
\textcolor{black}{and $\dot{M}$ is different from the mass-transfer 
rate from a companion star}.  
The value of $r_{\rm in}$ is set to 0.02 in units of the binary 
separation.  
The $\alpha$-value is also fixed to be 0.1.  
The critical value of $\dot{M}$ at a given radius, at which 
the disk returns in quiescence, depends on the radius in the 
accretion disk (equation (39) in \cite{ham98diskmodel}), and 
the mass accretion rate in outburst would become large 
in long orbital period systems.  
We set values of $\dot{M}$ as $2 \times 10^{-9}$, 
$3 \times 10^{-9}$, $4 \times 10^{-9}$, $5 \times 10^{-9}$, 
$6 \times 10^{-9}$ $M_{\solar}$ yr$^{-1}$ for 1SWASP J1621 
and BD Pav\footnote{If $\dot{M}$ is less than $1 \times 10^{-9}$ 
$M_{\solar}$ yr$^{-1}$, the outer disk cannot be in the outburst 
state (equation (39) in \cite{ham98diskmodel}).} and 
$6 \times 10^{-8}$, $8 \times 10^{-8}$, $1 \times 10^{-7}$ 
$M_{\solar}$ yr$^{-1}$ for V364 Lib, considering their orbital 
period.  
Also, we assumed the flat temperature distribution of 
a disk in quiescence (3,000 or 3,500 or 4,000 or 4,500K), 
referring to \citet{woo86zcha}.  

On the basis of the assumption of the steady state, we set 
the thickness of accretion disk in outburst as follows: 
\begin{eqnarray}
\frac{H}{r} = 1.72 \times 10^{-2} \alpha^{-1/10} \dot{M}_{16}^{3/20} {\frac{M}{M_{\solar}}}^{3/8} r_{10}^{1/8} \left[1-\left(\frac{r_{\rm in}}{r} \right)^{1/2} \right]^{3/5},  
\label{height-steady}
\end{eqnarray}
where $H$, $\dot{M}_{16}$ and $r_{10}$ represent the height of 
the disk, $\dot{M}/10^{16}$ g s$^{-1}$ and $r/10^{10}$ cm, 
respectively \citep{sha73BHbinary}.  
On the other hand, we assumed that the disk is flat in quiescence 
for simplicity.  The tested parameters of the height of the disk 
were 0.003, 0.005, 0.007, 0.009 in units of the binary separation.  

The masses of the primary and the companion are fixed.  
We used the values in Table \ref{property}.  
The radii of the white dwarfs in 1SWASP J1621, BD Pav, and 
V364 Lib are estimated to be 0.009, 0.008, and 0.008 $R_{\solar}$, 
respectively \citep{nau72WDmassradius}.  
We assumed that the temperature of the companion star is 4,300 K 
for 1SWASP J1621, 3,500 K for BD Pav, and 6,600 K for V364 Lib, 
considering the spectral types of the companions 
\citep{all73quantities} and the reasoning in Sec.~4.  
The temperature of the white dwarfs is fixed to 20,000 K.  
In addition, we need to set the distance from Earth to 
the object in order to determine the apparent magnitude.  

\begin{table*}
  \caption{Properties of 1SWASP J1621, BD Pav, and V364 Lib.  }
\label{property}
\begin{center}
\begin{tabular}{llcccc}
\hline
Object & ${P_{\rm orb}}^{*}$ & $M_{1}$$^{\dagger}$ & $q$$^{\ddagger}$ & Outburst History$^{\S}$ & Reference$^{\parallel}$\\
\hline
1SWASP J1621 & 0.207852(1) & 0.9 & 0.44 & 2006, 2016 & 1, 2, 3 \\
BD Pav & 0.17930 & 1.0 & 0.44 & 1938, 1985, 1996, 1997, 1998, 2000, 2006, 2015, 2017 & 4--10, This work \\
V364 Lib & 0.70243(11) & 1.0$\pm$0.2 & 1.5$\pm$0.3 & 2003, 2006, 2009, 2012, 2017 & 11--16, This work \\
\hline
\multicolumn{6}{l}{
$^{*}$Orbital period in units of d.} \\
\multicolumn{6}{l}{
$^{\dagger}$Mass of the primary white dwarf in units of $M_{\solar}$.} \\
\multicolumn{6}{l}{
$^{\ddagger}$Mass ratio of the companion star to the primary star ($q \equiv M_{2}/M_{1}$).} \\
\multicolumn{6}{l}{
$^{\S}$Years when outbursts were recorded.  } \\
\multicolumn{6}{l}{
\parbox{500pt}{$^{\parallel}$1: \citet{tho16j162117}, 2: \citet{dra14CRTSCVs}, 3: \citet{kju17j1621}, 4: \citet{sio08CVHST}, 5: \citet{fri90CVsodium}, 6: \citet{bar87bdpav}, 7: vsnet-alert 1008, 8: vsnet-alert 2388, 9: vsnet-alert 4888, 10: vsnet-alert 18762, 11: \citet{uem09asas1509atel}, 12: vsnet-alert 14271, 13: \citet{wil11j1714asas1509}, 14: \citet{poj09asas1509}, 15: \citet{kin09asa1509cbet}, 16: vsnet-alert 20877.}} \\
\end{tabular}
\end{center}
\end{table*}

\subsubsection{Limitation of Our Model and How to Determine the Best Model Parameters}

   Our modeling deals with only the emission from 
optically-thick regions and assumes a simple geometry, 
an axisymmetric disk, no hot spot and no hot corona.  
The hot spot is located near the disk edge, 
and an optically-thin hot corona expanding around the disk 
is produced by the evaporation of the inner disk at low 
accretion rate.  
These components, which are missing from our models, 
have high temperature ranging between $\sim$5,000--20,000 K 
\citep{liu95DNevaporation,sta01iyuma}.  
In quiescence and/or in high inclination systems, therefore, 
these optically-thin hot regions are likely dominant, and hence, 
our models have some difficulties reproducing the observations, 
especially in the $B$ band.  
In other words, the observations in outburst in the $R_{\rm C}$ 
band are expected to be reproduced more easily in our models 
than those at shorter wavelengths.  
In addition, we assume that the distribution of the disk 
temperature in outburst follows that in the steady state, 
but the real distribution may not.  
The lack of a hot spot, simple disk geometry, and simple 
distribution of the disk temperature in the models are 
also raised in \citet{zol17j1621} as the factors responsible 
for not completely explaining the observations.  

   On the basis of the limitations described in the previous 
paragraph, we gave priority to reproducing the $R_{\rm C}$-band 
phase-averaged profile when determining the best models of 
1SWASP J1621.  In addition, we note that the inclination must 
have the same value in outburst and quiescence.  
We thus chose the best models as for 1SWASP J1621 in the 
following way.  
\begin{enumerate}
\item
We estimated at first each $\chi^{2}$ value between 
the observational and calculated orbital phases in 
the outburst state with various tested parameters specified 
in Sec.~5.2.2, and determined the best model in that state 
by the minimum $\chi^{2}$ estimations in the $R_{\rm C}$ band.
\item
With the inclination fixed to its value in the 
best model in outburst, we estimated each $\chi^{2}$ 
value between the observational and calculated orbital phases 
in quiescence.  
We determined the best model by the minimum $\chi^{2}$ 
estimations in the $R_{\rm C}$ band.  
\end{enumerate}
   As for BD Pav and V364 Lib, we chose the best model 
in quiescence by the minimum $\chi^{2}$ values 
between the observed $V$-band phase-averaged profile and 
the calculated $V$-band one.  
In addition, we chose the best mass accretion rate, which 
can best reproduce the outburst amplitude in the $V$ band, 
with the disk radius and inclination fixed as the best-fit 
parameters in quiescence.  

It is difficult to employ sophisticated methods to calculate 
error bars due to the long computational times; therefore, 
we estimated the model profiles with the rough grids 
described in Sec.~5.2.2, and determined the 90\% confidence 
intervals as the range in which $\Delta \chi^2$ from 
the minimum $\chi^2$ is within 2.706, by computing 
the models in the finer grids than those described in 
Sec.~5.2.2.  
Since accurate constraints of the inclination and disk 
height in quiescence is beyond our current goals for 1SWASP 
J1621 and BD Pav, we fixed the inclination to the value 
which produces the minimum $\chi^2$ in these two objects.  
Additionally, it is difficult to restrict the ranges of 
the parameters which are related to the accretion disk 
in V364 Lib because of the bright companion; hence, 
we only provided an error bar for the inclination.

\subsection{Results}

\begin{figure}[htb]
\begin{minipage}{0.99\hsize}
\begin{center}
\FigureFile(70mm, 50mm){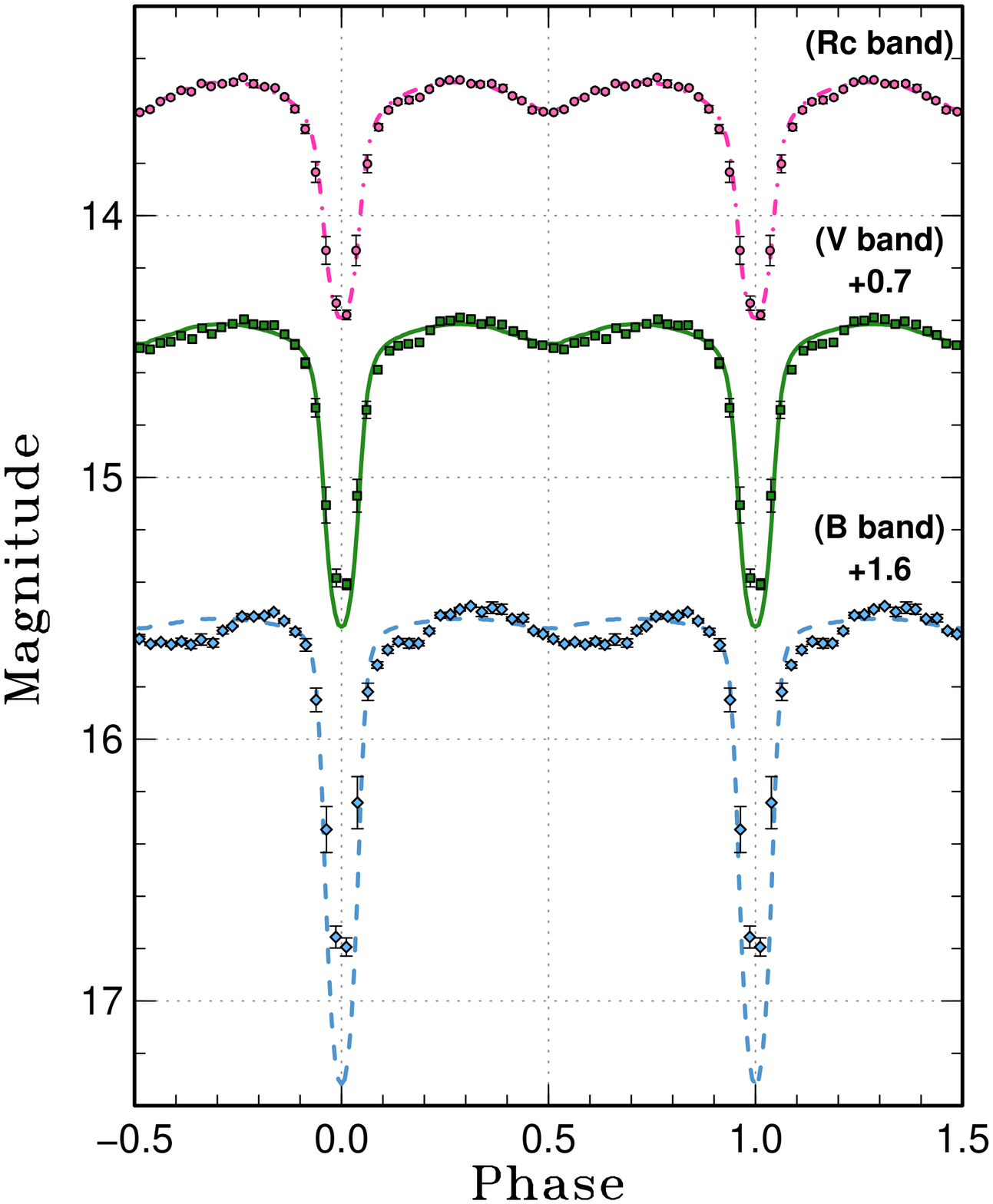}
\end{center}
\end{minipage}
\\
\begin{minipage}{0.99\hsize}
\begin{center}
\FigureFile(70mm, 50mm){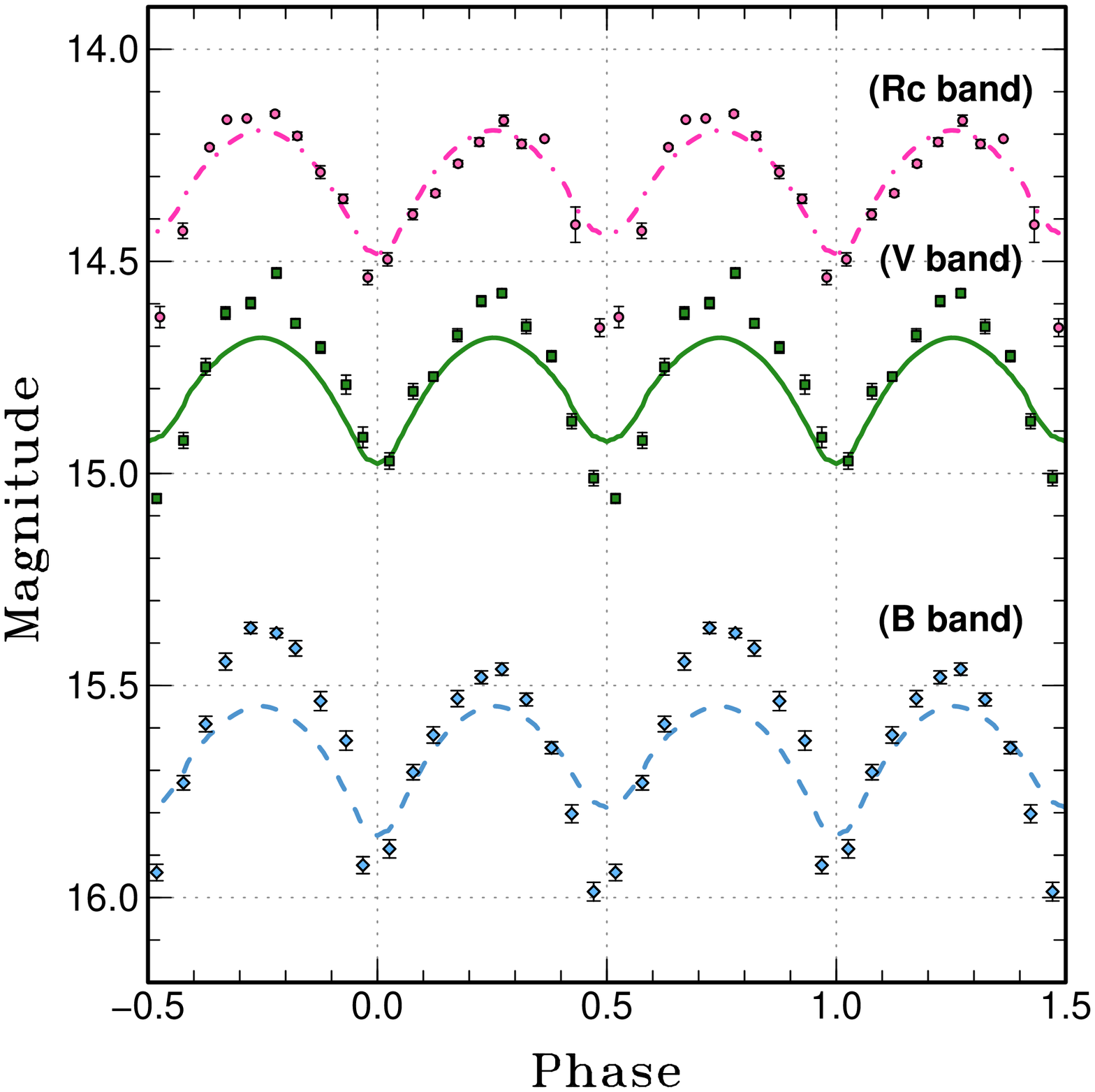}
\end{center}
\end{minipage}
\caption{Best models of the phase profiles of 1SWASP J1621 in outburst (upper panel) and quiescence (lower panel).  The dot-dash, solid and dash lines represent the calculated phase profiles in the $R_{\rm C}$, $V$, and $B$ bands.  The points with error bars are the observational phase profiles.  The magnitudes of the observational phase profiles are offset for visibility except for those in the $R_{\rm C}$-band ones in outburst.  }
\label{j1621model}
\end{figure}

   Our modeling suggests the inclination is close to 90 deg 
in 1SWASP J1621 (see Table \ref{parameter}).  The best-fit 
models are displayed in figure \ref{j1621model}.  
The very high inclination has been already noticed by 
\citet{kju17j1621} and \citet{zol17j1621}.  
Our models do not accurately fit the phase profiles 
in quiescence and in the $V$ and $B$ bands in outburst, 
but they do reproduce the main characteristics of 
quiescent double-waved modulations without a hot companion. 
The models in \citet{kju17j1621} and \citet{zol17j1621} 
for 1SWASP J1621 also did not reproduce the observations 
in all of the bands.  
Although our model includes more reasonable thicknesses 
of disks and temperature distributions of quiescent disks, 
compared to their models, it still had difficulty reproducing 
the quiescent profiles.  
This would be due to the limitations of our modeling, which 
are described in the first paragraph in Sec.~5.2.3.  
In particular, our model does not include any optically-thin 
region which is observable in high-inclination systems and 
in quiescence.  In addition, a hot spot would be dominant 
around the phase 0.75 in the $B$ band.  
Also, the calculated depth of the eclipse in the $B$ band in 
outburst is $\sim$0.5-mag deeper than the observed one.
Moreover, the calculated $B-V$ color is larger by $\sim$0.1 mag, 
and the calculated $V-R$ color was smaller by $\sim$0.15 mag 
than the observed one.  
Some fine-tuning of the disk temperature and structure may 
help to reproduce completely the observational depth 
of the eclipse and colors, but this is beyond our purpose.  
Since this system seems to have a very high inclination, 
there is a possibility that a flared disk structure may 
contribute to the emission.  
If the inclination is close to 90 deg, the hot white dwarf 
and the inner disk are hidden by the outer disk, 
as shown in the upper panel of figure \ref{configuration}.
In this case, the radiation from the hot white dwarf and 
the innermost region of the disk hardly contributes 
the overall emission from the object.  This situation will 
easily produce the shallow primary minima in quiescence.  

\begin{figure}[htb]
\begin{center}
\FigureFile(70mm, 50mm){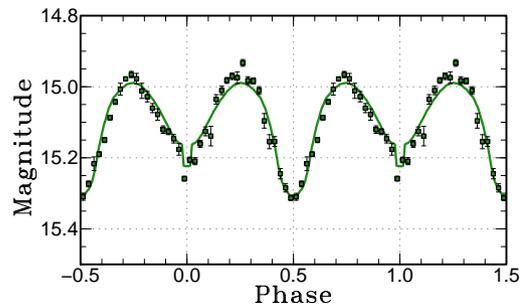}
\end{center}
\caption{Best model of the phase profile of BD Pav in quiescence.  The solid line represents the calculated phase profile in the $V$ band.  The points with error bars are the observational phase profile.  }
\label{bdpavmodel}
\end{figure}

   As for BD Pav, the inclination seems to be close to 
75 deg (see Table \ref{parameter}), which is consistent 
with previous studies \citep{sio08CVHST,fri90CVsodium} 
and the estimation in Sec.~5.1. 
Since our modeling does not explain the observations 
well enough to estimate error bars, we just provide 
the best model parameter in the rough grids.  
The reasons why our model does not completely reproduce 
the observations are the limitations described 
in Sec.~5.2.3; in addition, our sparse profile may make 
it difficult to recognize the eclipse of the white dwarf 
in this object.  
The best-fit model in quiescence which is given in figure 
\ref{bdpavmodel}, however, reproduces the W UMa-like orbital 
modulations without a hot companion as in 1SWASP J1621.  
The mass accretion rate to reproduce the outburst amplitude 
is in the range expected by the disk-instability model.  
In addition, according to \citet{bar87bdpav}, the $B-V$ 
colors in the outburst and quiescent states are $\sim$0.1 
and $\sim$0.6 mag, and the $V-R$ colors in the outburst and 
quiescent states are $\sim$0.0 and $\sim$0.5 mag, 
respectively; our modeling reproduces well the $V-R$ 
colors.  
Although the calculated $B-V$ color in quiescence was 
$\sim$0.6 mag larger than the observational one, the reported 
colors in \citet{bar87bdpav} were measured at the end of 
the outburst.  
Some time after the outburst, the $B-V$ color might become 
large as the disk becomes cool.  
Though the grazing eclipses in the systems having 
$\sim$70-deg inclinations are characterized by deep primary 
minima in quiescence as in U Gem (chapter 2.4.3 in 
\cite{hel01book}) due to bright hot spots, the eclipse in 
BD Pav does not agree with this behavior.  This would be 
because of the dim hot spot as suggested in Sec.~5.1.  
If the inclination is 75 deg, the white dwarf is barely 
eclipsed (see also the middle panel of figure 
\ref{configuration}).

\begin{figure}[htb]
\begin{center}
\FigureFile(70mm, 50mm){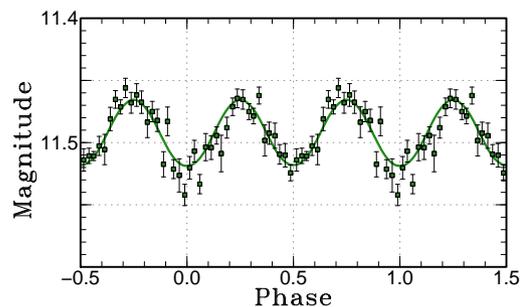}
\end{center}
\caption{Best model of the phase profile of V364 Lib in quiescence.  The solid line represents the calculated phase profile in the $V$ band.  The points with error bars are the observational phase profile.  }
\label{asas1509model}
\end{figure}

\begin{figure*}[htb]
\begin{center}
\FigureFile(160mm, 100mm){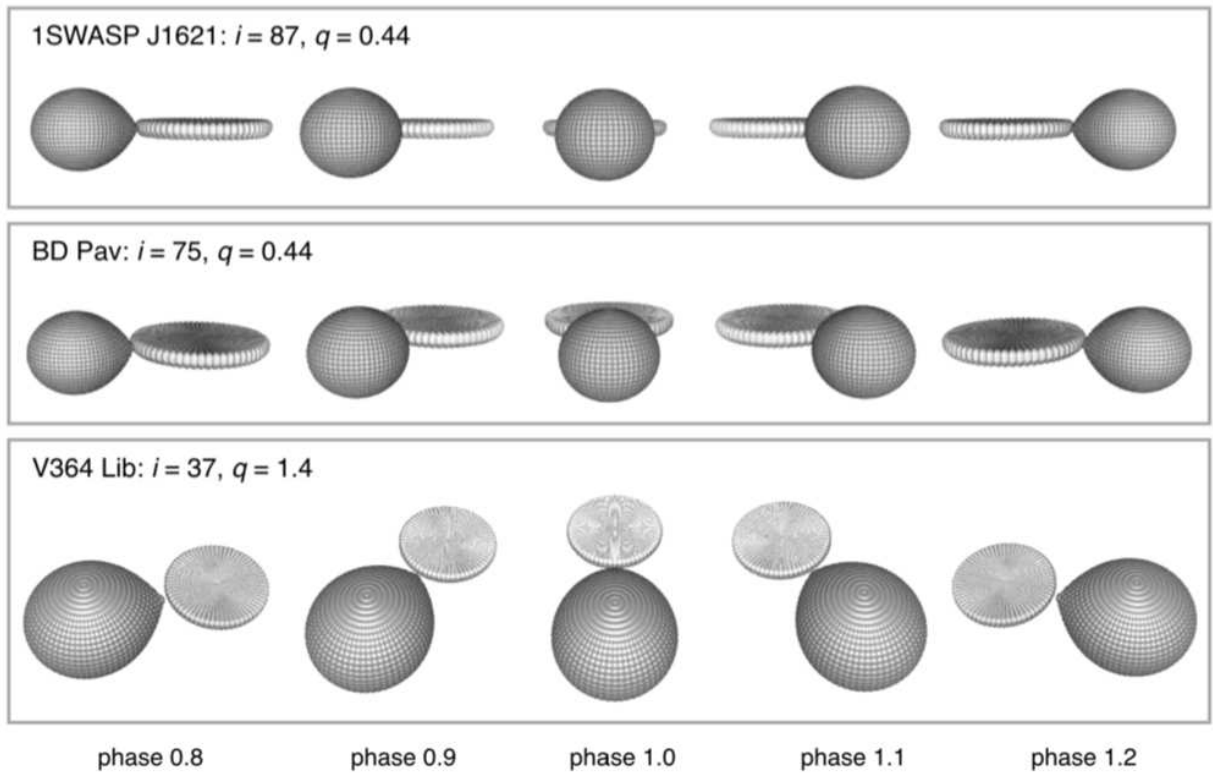}
\end{center}
\caption{Model configurations around the eclipses of the accretion disks in 1SWASP J1621, BD Pav, and V364 Lib in the outburst state.  }
\label{configuration}
\end{figure*}

   As for V364 Lib, we confirmed that the ellipsoidal 
variations of the hot companion are dominant in quiescence, 
and that the inclination is low (see Table \ref{parameter}).  
The estimated inclination is consistent with that in 
Sec.~4.  The best-fit model is shown in figure 
\ref{asas1509model}.  
The accretion disk is not eclipsed, but the ellipsoidal 
variations are observable due to the hot companion star 
(see the lower panel of figure \ref{configuration}).  
As for the outburst state, \textcolor{black}{the value of 
$\dot{M}$ is chosen to reproduce the outburst amplitude.  
The $\dot{M}$ value much larger than those in the other 
two objects seem to originate from its long orbital 
period.  
The critical value of $\dot{M}$ at which the transition 
from the hot branch to the cool branch on the thermal 
equilibrium curve is triggered, is approximately 
proportional to $r^{2.7}$ 
\citep{ham98diskmodel}\footnote{\textcolor{black}{Here, 
$r$ is the same one as defined in Sec.~5.2.2.}}.  
If the hot state extends to the outer disk in its outbursts, 
the mass accretion rate in outburst is likely much higher 
than those in shorter-period dwarf novae, in spite of our 
rough estimations.}
We also confirmed very weak ellipsoidal variations with 
amplitudes of less than 0.03 mag via the model 
with the parameters given in Table \ref{parameter}; 
this is consistent with our observations.  
The reason for the lack of prominent ellipsoidal variations 
in outburst is that the disk becomes much brighter 
in the outburst than that in quiescence.  
We consider that the bright disk in outburst reduces the relative 
contribution of the ellipsoidal variations of its companion 
star, and that some variations hidden by the bright 
companion during quiescence become visible in outburst.   
In addition, the outburst amplitude is reproduced with 
the parameter in Table \ref{parameter}.  
The colors of $V-R$ and $B-V$ in quiescence are reported 
to be 0.1 and 0.5 mag, respectively, by \citet{APASS}, and 
our calculated colors were almost identical.  
In the systems having a high-temperature companion, its 
contribution dominates the whole emission from the dwarf 
nova.  If these systems have high inclinations, the eclipses 
of companions may become deeper than those of the disks and 
white dwarfs in quiescence, and the eclipses of disks and 
white dwarfs are prominent in outburst.  These are 
inconsistent with the observations of V364 Lib.  

\begin{table}
	\caption{Model parameters to reproduce the eclipsing light variations and computed outburst amplitudes and colors in 1SWASP J1621, BD Pav, and V364 Lib.  }
	\label{parameter}
	\centering
	\begin{tabular}{cccc}
\hline
Parameters & 1SWASP J1621 & BD Pav & V364 Lib \\
\hline
${i}^{*}$ & 87 & 75 & 37$^{+1}_{-4}$ \\
${R_{\rm disk}^{\rm burst}}^{\dagger}$ & 0.89$^{+0.01}_{-0.04}$ & 0.90 (fix) & 0.80 (fix) \\
$\dot{M}^{\ddagger}$ & 5.4$^{+0.1}_{-0.3}$$\times 10^{-9}$ & 2$\times 10^{-9}$ & 1$\times 10^{-7}$ \\
${R_{\rm disk}^{\rm qui}}^{\S}$ & 0.86$^{+0.04}_{-0.01}$ & 0.90 & 0.80 \\
${T_{\rm disk}}^{\parallel}$ & 4,470$^{+30}_{-70}$ & 3,000 & 4,500 \\
${h_{\rm disk}}^{\#}$ & 0.009 & 0.007 & 0.003 \\
$d^{\P}$ & 166 & 253 & 536 \\
\hline
Amp$^{**}$ & 0.70 & 3.11 & 0.85 \\
${{V-R}_{\rm out}}^{\dagger\dagger}$ & 0.23 & 0.10 & 0.06 \\
${{B-V}_{\rm out}}^{\ddagger\ddagger}$ & 0.23 & 0.04 & 0.05 \\
${{V-R}_{\rm qui}}^{\S\S}$ & 0.49 & 0.73 & 0.20 \\
${{B-V}_{\rm qui}}^{\parallel\parallel}$ & 0.87 & 1.12 & 0.36 \\
\hline
\multicolumn{4}{l}{$^{*}$Inclination angle in units of deg.}\\
\multicolumn{4}{l}{$^{\dagger}$Disk size in the outburst state in units of $R_{\rm L1}$.}\\
\multicolumn{4}{l}{\parbox{230pt}{$^{\ddagger}$Mass accretion rate in outburst \textcolor{black}{in units of $M_{\solar}$ yr$^{-1}$.  We assume the steady state in outburst.}}}\\
\multicolumn{4}{l}{$^{\S}$Disk size in quiescence in units of $R_{\rm L1}$.}\\
\multicolumn{4}{l}{$^{\parallel}$Temperature of a disk in quiescence in Kelvin.}\\
\multicolumn{4}{l}{$^{\#}$Thickness of a disk in units of its binary separation.}\\
\multicolumn{4}{l}{\parbox{230pt}{$^{\P}$Distance from Earth to the object, which reproduces the apparent magnitude in the $R_{\rm C}$ band in outburst as for 1SWASP J1621, and in the $V$ band in quiescence as for BD Pav and V364 Lib, in units of pc.  }}\\
\multicolumn{4}{l}{\parbox{230pt}{$^{**}$Outburst amplitude in the $R_{\rm C}$ band at phase 0.25 in units of magnitude.  It is derived from the best models in outburst and quiescence.}}\\
\multicolumn{4}{l}{\parbox{230pt}{$^{\dagger\dagger}$$V-R$ color in outburst at phase 0.25 in units of magnitude.}}\\
\multicolumn{4}{l}{\parbox{230pt}{$^{\ddagger\ddagger}$$B-V$ color in outburst at phase 0.25 in units of magnitude.}}\\
\multicolumn{4}{l}{\parbox{230pt}{$^{\S\S}$$V-R$ color in quiescence at phase 0.25 in units of magnitude.}}\\
\multicolumn{4}{l}{\parbox{230pt}{$^{\parallel\parallel}$$B-V$ color in quiescence at phase 0.25 in units of magnitude.}}
\end{tabular}
\end{table}

\section{Discussion}

\subsection{Nature of the Three Objects and Their Low-Amplitude Outbursts}

Although 1SWASP J1621, BD Pav, and V364 Lib show similar 
outburst behavior, their natures are not necessary the same.  
As mentioned in the introduction, some people consider 
1SWASP J1621 to have \textcolor{black}{an unusually hotter 
companion star than normal K or M-type companions in dwarf 
novae (see Figure 2.45 in \cite{war95book})} according to 
\citet{waa17j1621}, 
but our results show that the characteristics of the orbital 
phase profiles of 1SWASP J1621 can be explained by its very 
high inclination and faint hot spot, without a hot companion 
star.  This condition is similar to that in BD Pav.  
On the other hand, for V364 Lib, our results support that 
this system has a low inclination and 
\textcolor{black}{an F-type hot companion star, which is expected 
to be bright (large) by the long orbital period and the large 
mass ratio (see also Sec.~4).}

Their properties are likely relevant to their small-amplitude 
($\sim$1--2 mag) outbursts (see Sec.~3).   
In the cases of 1SWASP J1621 and BD Pav, the critical 
conditions are their high inclinations and inside-out 
outbursts.  
In high-inclination systems, the observable disk surface 
is small.  In addition, if a system has a very high 
inclination, as in 1SWASP J1621, the hot inner region of 
the accretion disk is hidden by the outer disk located 
along the line of sight.  Although the inner disk becomes 
bright in the outburst state, it cannot be observed.  
The inside-out outbursts expected from the slow-rise 
shapes of the outbursts (see Sec.~3) 
would be another key factor of the low-amplitude outbursts.  
The outburst amplitudes of inside-out outbursts are often 
smaller than those of outside-in outbursts, since the slowly 
propagating heating front prevents the whole disk from 
entering the hot state \citep{sma84DI,can86DNburst}.  
On the other hand, in V364 Lib, its hot companion star 
is the main factor in its low-amplitude outbursts, though 
inside-out outbursts may also contribute to the small 
outburst amplitudes.  If the companion is unusually hot and big, 
its contribution to the whole brightness is large at optical 
wavelengths.  The change of the disk brightness 
between the quiescence and outburst states would be 
unremarkable at optical wavelengths even if the total 
luminosity emitted from the disk is high in outburst.  
We suggest that astronomical surveys might easily overlook 
such a low-amplitude outburst.

\subsection{Mass-Transfer Bursts ?}

We believe that the outbursts in the three objects are 
explained well by the disk-instability model.  
\citet{qia17j1621} claimed the mass transfer ceases 
because of the star spot on the surface of the companion star, 
which is located at the $L_{1}$ point in quiescence, and the 
outbursts are caused by an abrupt mass transfer because of 
the expanding companion in 1SWASP J1621.  However, our results 
show the existence of the large accretion disk in the outburst 
state in this object, and this is inconsistent with their argument.  
If a lot of matter is transferred to the primary side at the 
onset of the outburst, the radius of the accretion disk 
becomes close to the Lubow-Shu radius \citep{lub75AD}, which 
is approximately expressed by $0.0859 q^{-0.426}$ in units 
of the binary separation \citep{hes90gd552}, because of 
the conservation of angular momentum \citep{ich92diskradius}.  
This value is much smaller than the estimated disk size in our 
modeling.  When a small disk is occulted by the companion, 
the shape of the bottom of primary minima becomes flat, and 
this phenomenon disagrees with the morphology of the 
phase profiles observed in the outburst state.  
In addition, continuous accretion during quiescence may 
help to build a large disk in outburst by removing matter 
with small specific angular momentum.  
\citet{zol17j1621} also concluded that their modeling of 
the eclipsing profile in quiescence suggests a white dwarf 
plus a large accretion disk, and it is consistent with our 
results in Sec.~5.3.  
Their results, and ours, argue against the halting of 
mass transfer in quiescence, which is a key element in 
\citet{qia17j1621}.  
Therefore, there is no need for introducing mass transfer bursts 
to explain the outbursts in this kind of object.  
\citet{qia17j1621} proposed their interpretation based in part 
on the observations of \citet{pav16j1621}, 
but the short-term periodic variations reported in this 
reference were not confirmed in our observations.  

Actually, there are some contradictions between the observations 
and the discussion in \citet{qia17j1621}.  For example, they considered 
that only the eclipse of the white dwarf was observed in quiescence, 
but the width of the eclipse that they regard as the ingress of 
the white dwarf, about 0.007 d, indicates that the eclipsed 
object must have a width of more than 10 times of the radius 
of a 0.9-$M_{\solar}$ white dwarf.\footnote{The width 
of the ingress of the white dwarf is derived from the data 
in figure 2 in \citet{qia17j1621}.}  
In addition, the depth of the eclipse, about 0.2 mag, is 
not consistent with the eclipse of a white dwarf.  
If the temperature of the white dwarf is 20,000 K, the 
contribution of the white dwarf to the brightness of 
the entire system is about 10 \%.  The depth would then be 
less than 0.1 mag in their observations.  Since they consider 
that accretion onto the white dwarf is stopped during quiescence, 
the white dwarf should have a low-temperature in their discussion.  
If the temperature of the white dwarf is only 10,000 K, the depth 
of eclipses would be less than 0.01 mag.  
Moreover, they insisted no lasting disk in this object.  
If the contribution of the disk is very weak and the disk 
size is small in outburst, the flux at the primary minima 
in outburst would be the same as those in quiescence; however, 
this is inconsistent with figure 3 in \citet{qia17j1621}.

\subsection{Infrequent, Long-lasting and Inside-out Outbursts}

   Though we may overlook the outbursts due to their small 
amplitudes, the frequency of the outbursts in these systems 
is relatively low in comparison with many other dwarf novae 
above the period gap (see also Table \ref{property}), 
\textcolor{black}{
and their outburst durations are relatively long.  
We discuss below the origin of the infrequent, 
long-lasting, and inside-out outbursts.}

\textcolor{black}{
   As for the origin of the infrequent outbursts in 1SWASP 
J1621 and BD Pav, we consider that it is attributed to low 
mass-transfer rates.}  This is because 1SWASP J1621 and BD Pav 
have clearly longer outburst intervals than many other 
dwarf novae having 4--5 hours orbital periods.  
In addition, the almost symmetrical profiles with respect 
to the primary minima in these two objects (see figures 
\ref{phase} and \ref{bdpav-phase}) suggest that the hot spots 
are faint.  This implies that the mass transfer rate is low, 
since the brightness of a hot spot depends on it.  
\textcolor{black}{  
   On the other hand, the origin of the infrequent outbursts 
in V364 Lib would not be necessarily low mass transfer rate.}  
We suspect its long orbital period is related to the origin.  
This is because it would take much time to form a dense disk 
enough to trigger outbursts in V364 Lib due to its long orbital 
period which means a large value of $R_{\rm L1}$.  
The relation between \textcolor{black}{long outburst intervals} and 
long orbital periods in several dwarf novae has been 
pointed out in \citet{bru17v1129cen}, and we show 
it in figure \ref{porbint}.\footnote{The values of 
outburst intervals in the long-period objects were derived from 
\citet{men86bvcen,she09v630cas,sek92v1017sgr,bru17v1129cen} 
and see also figure E1 in the supplementary information of 
this paper.  
V1017 Sgr also has a very long orbital period, but we excluded it 
from the regression shown in figure \ref{porbint} because it is 
clearly out of the trend.  }
For instance, compared with an typical dwarf nova having 
an orbital period of 4 hours, a mass ratio of 0.45 and 
outburst intervals of 30--100 days, V364 Lib has 
an $R_{\rm L1}$-value about 2.6 times larger. 
If the transfer rate is almost equal to that in a typical 
dwarf nova, 
\textcolor{black}{then the accretion disk in V364 Lib would accumulate 
$\sim$2.5--7-times more mass than that in a typical dwarf nova 
until an outburst is triggered.  }
In other words, based on mass transfer in 
ordinary dwarf novae, we would expect V364 Lib to have 
an outburst interval of about 70--700 days, but this value 
seems to be shorter than the observational outburst intervals 
in V364 Lib.  Actually, V364 Lib may not follow the linear 
relation between the orbital period and the logarithm of 
the outburst intervals in these systems, 
\textcolor{black}{although all 
outbursts in this object might not be detected due to their 
small amplitudes.}  
Since the above discussion is a simple approach, our estimation 
has some uncertainty.\footnote{\textcolor{black}{For example, 
the mass transfer rate is likely larger in the systems having 
longer orbital periods} so that the rate in this object might be 
about 4 times larger than that in an typical dwarf nova 
according to equations (38) and (39) in \citet{ham98diskmodel}.  
It is, however, possibly similar to the typical one, because 
this system enters inside-out outbursts.  In addition, 
the structure and density profile in the disk would also affect 
the outburst intervals, and the expected intervals may become 
longer.  }
\textcolor{black}{We thus could not reject either of the long 
orbital period or the low mass-transfer rate as the origin 
of the infrequent outbursts in V364 Lib.  The value of $\dot{M}$ 
in Table \ref{parameter} represents not the mass-transfer rate 
but the accretion rate in outburst.  
The actual mass transfer rate should be smaller than that 
value.  }

\begin{figure}[htb]
\begin{center}
\FigureFile(80mm, 50mm){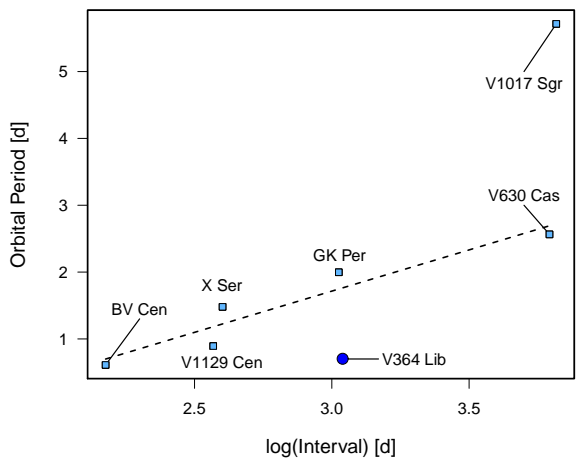}
\end{center}
\caption{Linear relation between the orbital periods and the logarithm of the outburst intervals in several dwarf novae having very long orbital periods.  The dashed line represents the regression formula $y = 1.2 x - 2.0$ estimated by the least-squares method with the data of BV Cen, V1129 Cen, GK Per, X Ser and V630 Cas.  Here, $x$ and $y$ represent the orbital period and the interval between outbursts.  }
\label{porbint}
\end{figure}

The inherent low-mass transfer rates and/or long orbital 
periods would naturally explain \textcolor{black}{the characteristic 
long outburst durations and inside-out outbursts} 
in the three objects within the context of the disk 
instability model.  
\textcolor{black}{
This is because a lot of mass is expected to accumulate 
in the disk and because the mass would be easily diffused 
inward on viscous timescales in these conditions.}  
Actually, the numerical simulations of the disk instability 
model showed that, in the dwarf nova GK Per having a very long 
orbital period, most of its outbursts are easily triggered 
at the inner disk, and that this behavior cannot be reproduced 
by mass-transfer bursts \citep{kim92gkper}.  
It is also consistent to the idea of the disk-instability 
model that the large stored disk mass before the onset 
of an outburst can be related to the long outburst durations.  
The slow rises of these outbursts, in addition to their small 
amplitudes, may also make it difficult to detect them.  
Long-lasting inside-out outbursts were commonly observed 
in V1129 Cen and HS 0218, which are similar to V364 Lib and 
1SWASP J1621, respectively \citep{bru17v1129cen,gol12hs0218}; 
\citet{gol12hs0218} have already pointed out that 
an inside-out outburst is expected from systems with low 
mass-transfer rates, which is considered to be associated 
with the almost symmetrical phase profiles, as in HS 0218.  

   \textcolor{black}{The mechanism that causes lower mass-transfer 
rates in the three objects than those in many other dwarf 
novae above the period gap may be unclear.}  
One plausible explanation is that the mass-transfer rate varies 
in the same system on much longer timescales than the interval 
between outbursts for some reason.  
For example, \citet{sha86hibernation} and \citet{kov88hibernation} 
proposed hibernation scenario.  According to their proposition, 
during about a century after a nova eruption, the mass-transfer 
rate remains high due to the strong irradiation from 
the white dwarf, and thereafter, it decreases.  
V1213 Cen, a classical nova, revealed dwarf-nova 
outbursts 6 years before its eruption \citep{mro16v1213cen}, 
and this phenomenon is understood as evidence of hibernation.  
The hibernation is, however, believed to occur 
only when the binary separation expands just after a nova 
eruption.  On the basis of equations (14) and (21), 
which represent the relation between the variation of 
binary separation and the mass ratio in 
\citet{sha86hibernation}, it is doubtful whether the binary 
separation expands after a nova eruption in V364 Lib.  
1SWASP J1621, BD Pav may be in hibernation, but V364 Lib 
may not.

\subsection{Appearance of Highly Ionized Emission Lines}

   The He II 4686 emission line was observed in 1SWASP J1621, 
BD Pav, and V364 Lib during the outburst state (\cite{sca16j162117}; 
\cite{bar87bdpav}; Sec.~4).  In addition, the C III/N III emission 
line was also observed in the outburst of V364 Lib.  These 
highly ionized lines are barely detected in the outbursts of many 
dwarf novae.  
In the case of V364 Lib, a massive white dwarf is a natural 
consequence on the basis of its massive companion star (see Sec.~4).  
Since the primary star has degenerated earlier, 
the progenitor of the primary white dwarf would be more massive 
than a F-type companion star as for V364 Lib.  For example, 
the relation between the final white-dwarf mass and the initial 
main-sequence mass given in Fig.~2 in \citet{sal09WDmass} shows 
the white-dwarf mass can exceed 1$M_{\solar}$ if the initial 
mass exceeds $\sim$6$M_{\solar}$, and our observations do not 
deny the possibility that the initial mass of the primary 
exceeds $\sim$6$M_{\solar}$.  
It is thus suggested that the primary white dwarf in this system 
is likely heavier than those in the CVs which do not have 
massive donor stars.  In spite of the large uncertainty, 
our estimation in Sec.~4 implies the possibility that this 
object has a massive (more than 1$M_{\solar}$) white dwarf.
If the white dwarf is massive, the inner disk has a relatively 
high temperature.  The massive white dwarf may be, therefore, 
the origin of the highly ionized emission lines in this 
system.  
We note that some dwarf novae having massive white dwarfs 
showed highly-ionized emission lines in outburst 
(e.g., \citet{wil15gkper,war88bvcen,mor02DNspectralatlas}).  
In addition, this type of line was detected in a recurrent 
nova having a massive white dwarf \citep{dia10usco}.

   On the other hand, the source to produce the He II emission 
line in outburst in normal dwarf novae having $\sim$0.8-$M_{\solar}$ 
white dwarfs \citep{zor11SDSSCVWDmass} is unclear, but such 
a highly ionized emission line is sometimes observed in high 
inclination systems (e.g., \cite{bab02wzsgeletter,har92zchaUVspec}).  
The He II emission line in 1SWASP J1621 and BD Pav may arise 
due to their high inclinations, although their relatively massive 
white dwarfs, heavier than $\sim$0.8-$M_{\solar}$, may contribute 
to the line as in V364 Lib.  
This possibility is also notified by \citet{zol17j1621}.  
In the high-inclination systems, the accretion disk is optically 
thick.  Additionally, the He II emission line in the 1985 outburst 
of BD Pav showed a single-peaked profile according to figure 1 in 
\citet{bar87bdpav}, although our modeling and other works 
suggest the inclination of this system is $\sim$75 deg.  
The emission line from the accretion disk will have a clear 
double-peaked shape due to its high inclination 
\citep{hor86CVlineformation}.  
Thus the radiation is not likely to originate from a disk.  
Such a single-peaked He II emission line was also observed in 
outburst in HT Cas, another dwarf nova, and a disk wind is 
proposed as its possible origin \citep{mur97diskwind}.  
In the outbursts of dwarf novae, the disk temperature becomes 
high as in nova-like systems.  The single-peaked emission 
lines whose origin is interpreted as the disk wind are often 
observed in nova-like systems (e.g., \cite{bap02ippeg}), 
and hence, it would not be strange that the strong single-peaked 
emission lines are detected in some dwarf-nova outbursts.  
Additionally, an optically-thin disk wind may contribute 
to the phase profile in the $B$ band.

\subsection{Evolutionary Stage of V364 Lib}

   It seems to be strange at first glance that V364 Lib, 
whose mass ratio is likely more than 1 has low mass transfer 
rate.  
In a subclass of CVs which have a companion more massive than
the primary white dwarf, if the companion star completely 
fills with its Roche lobe, the mass transfer becomes easily 
unstable since the mass loss from the companion star induces 
the binary separation to shrink.  We call this subclass 
``super soft sources''.  
In these systems, the mass transfer rate is as high as 
1--4$\times$10$^{-7}$ $M_{\solar}$ yr$^{-1}$ (see 
\cite{kah97SSSreview} for a review and references therein).  
This value is more than 100 times higher than the typical 
mass transfer rate in dwarf novae \citep{sha86CVmasstransfer}.   
On the contrary, if we consider the whole amount of the stellar 
wind flows via the $L_{1}$ point under the condition that 
the radius of the companion star has not yet filled with its 
Roche lobe, the mass transfer rate is too low to trigger 
dwarf-novae type outbursts.  
Under the assumption that the wind velocity is the escape velocity, 
the mass transfer rate estimated to be $\sim$$10^{-11-12}$ 
$M_{\solar}$/yr by using the equation $\dot{M} = 4\pi {R_2}^2 
\rho v_{\rm wind}$. 
We finally suggest that in V364 Lib, the thin outer layer of 
the companion fills with its Roche lobe.  Since the estimated 
value of the Roche radius on the companion side is about twice 
of the radius of a typical main-sequence F-type star, 
the companion is likely a sub-giant if our suggestion is 
correct.  

   Although the mass transfer rate in V364 Lib seems to be low 
at present, it may increase as the companion evolves and/or 
the orbital angular momentum decreases.  
Then this system may show nova eruptions repeatedly since the white 
dwarf is likely massive \citep{fuj82nova2}.  
In addition, if the mass transfer becomes unstable (i.e., the 
transfer rate becomes close to $\sim 10^{-7} M_{\solar}/{\rm yr}$), 
the system behaves as a super soft source.  
V364 Lib might eventually become a supernova via the super-soft-source 
phase.  
The duration of the post-common-envelope phase, before the system 
evolves to a completely semi-detached binary, is likely long, 
since the companion star evolves on timescales of $\sim$10$^{9}$ yr 
and the orbital angular momentum is decreasing on timescales of 
$\sim$10$^{8-9}$ yr (e.g., figure 12 in \cite{kni11CVdonor}).  
It would not be strange that some CV-like systems having 
massive companions show dwarf-nova-like outbursts.  
Recent numerical simulations on the CV evolution suggest 
that some CVs having F-type donor stars enter dwarf-nova type 
outbursts (figures 4 and 16 in \cite{kal16evolution}).

\section{Conclusions}

We presented the optical photometric observations of 1SWASP J1621, 
BD Pav, and V364 Lib and the optical spectroscopy of V364 Lib, 
and investigated their properties by using the numerical modeling.  
Our main findings and discussion are summarized in the following 
descriptions.  

\begin{itemize}
\item
There are many common features in the outburst behavior 
and the orbital variability in 1SWASP J1621, BD Pav, and 
V364 Lib.  
They show infrequent and small-amplitude outbursts.  
The outburst durations are long (about a few tens of days), 
and slow rises in brightness are observed at an early stage.  
In addition, all of the three systems show prominent 
ellipsoidal variations in quiescence.  
\item
In spite of the similarities in the outburst behavior, 
some properties in each system are somewhat different from 
the others.  
Our observations and numerical modeling suggest that 
1SWASP J1621 has a very high inclination ($\sim$90 deg) 
plus a faint hot spot, and these properties are similar to 
those in BD Pav, though it has a little lower inclination 
($\sim$75 deg).  
On the other hand, V364 Lib likely has an F-type bright 
companion star, a massive white dwarf, and a low inclination 
angle.  
\item
The characteristic low-amplitude outbursts in the three 
objects can be explained by a high inclination or a hot 
companion and inside-out outbursts.   
\item
The almost symmetrical phase profiles with respect to 
the primary minima, which arise due to faint hot spots, and 
long intervals between outbursts, would suggest a low 
mass-transfer rate in 1SWASP J1621 and BD Pav.  
The long orbital period implies the large disk for V364 Lib.  
The low mass-transfer rate and/or the large disk may be 
related to long outburst intervals, long outburst durations, 
and inside-out outbursts in these three objects.  
\item
The commonly observed, highly ionized emission lines in their 
outbursts would originate from the high inclinations, for 
1SWASP J1621 and BD Pav, and from the massive white dwarf for 
V364 Lib.  
A disk wind may be related to the profile of these highly 
ionized emission lines.  
\item
In spite of a donor star more massive than the primary 
white dwarf, V364 Lib shows dwarf-nova like outbursts.  
This implies only the thin outer layer of the donor star 
fills its Roche lobe in this system.  This object may 
evolve into a recurrent nova or a super soft source.  
\end{itemize}

Although the properties of the three objects are somewhat 
different from normal dwarf novae, (in particular, 
V364 Lib is different from many other dwarf novae having 
cool companions), 
our results show that the peculiar outburst behavior and 
the changing orbital profiles between outburst and 
quiescence in the three objects can be explained in terms 
of the disk instability model in normal dwarf novae.  
It is not necessary to consider strong magnetic activity 
and mass transfer bursts.  
The estimation of the mass ratio in V364 Lib has a large 
error, and hence, high-dispersion spectroscopy in the 
outburst state is important to determine it with high accuracy.  
We consider that the three systems recently reported by 
\citet{bru17gyhya,bru17v1129cen,gol13hs0218} have properties 
similar to those of 1SWASP J1621, BD Pav, or V364 Lib, 
and our discussion is applicable to their peculiar outburst 
behavior.  
Although the total number of this kind of object may be large, 
it is difficult to find them due to their low-amplitude and 
rare outbursts.

\section*{Acknowledgements}

This work was financially supported by the Grant-in-Aid for JSPS 
Fellows for young researchers (M.~Kimura, R.~Ishioka) and 
the Grant-in-Aid 
from the Ministry of Education, Culture, Sports, Science and 
Technology (MEXT) of Japan (25120007, T.~Kato and 17917351, D.~Nogami). 
We appreciate All-Sky Automated Survey for Supernovae (ASAS-SN) 
detecting a large amount of DNe.  
We are thankful to many amateur observers for providing a lot of 
data used in this research.
The numerical code that we used in this paper is provided by 
Izumi Hachisu.  
This work was partially supported by Grants VEGA 2/0008/17, 
APVV-15-0458 (S.~Shugarov), NSh 9670.2016.2 (N.~Katysheva, 
S.~Shugarov), RSF-14-12-00146 (P.~Golysheva, for processing 
observations data). O.~Vozyakova and S.~Shugarov acknowledges 
(partial) support from M.~V.~Lomonosov Moscow State 
University Program of Development.

\section*{Supporting information}

Additional supporting information can be found in the online version 
of this article:
Supplementary tables E1, E2, E3, E4, E5, E6, and E7 and figure E1.

\newcommand{\noop}[1]{}


\renewcommand{\thetable}{%
  E\arabic{table}}
\renewcommand{\thefigure}{%
  E\arabic{figure}}

\addtocounter{table}{-2}
\begin{table*}
	\caption{List of Instruments.}
	\label{telescope}
	\centering
	\begin{tabular}{lccc}
		\hline
		CODE$^{*}$ & Telescope (\& CCD) & Observatory (or Observer) & Site\\
		\hline
BSM & 25cmSC+Moravian G2-1600 & Flarestar Observatory & San Gwann, Malta \\
COO & PF 74 cm reflector+SBIG STL1001E & Lewis Cook & California, USA \\
CRI & 38cm K-380+Apogee E47 & Crimean astrophysical observatory & Crimea \\
GFB & CDK 50cm+Apogee U6 & William Goff & California, USA \\
IMi & 35cmSC+SXVR-H16 & Ian Miller & Furzehill Observatory, UK \\
Ioh & 30cmSC+ST-9XE CCD & Hiroshi Itoh & Tokyo, Japan \\
Kis & 25cm SC+Alta F47 & Seiichiro Kiyota & Kamagaya, Japan \\
Mhh & 10.5cm+SBIG ST-8XME & Hiroyuki Maehara & Japan \\
Mic & 84cm+E2V 42-40 & San Pedro Martir Observatory & Baja California, Mexico \\
MLF & 30cmRCX400+ST8-XME & Berto Monard Calitzdorp & South Africa \\
    & 35cmRCX400+ST8-XME & Berto Monard Calitzdorp & South Africa \\
OKU & 51cm+Andor DW936N-BV & OKU Astronomical Observatory & Osaka, Japan \\
RPc & FTN 2.0m+E2V 42-40 & LCOGT$^{\ddagger}$ & Hawaii, USA \\
    & 35cmSC+SXV-H9 CCD & Roger D. Pickard & UK \\
RIT & 30cm+ST-9E & RIT Observatory & New York, USA \\
SGE & 43cmCDK+STXL-11002 & Sierra Remote Observatories & Auberry, CA, USA \\
SHU & 18/225cm Maksutov telescope+SBIG ST-10XME & Stara Lesna & Slovakia \\
    & 60/750cm Zeiss telescope+FLI ML 3041 & Stara Lesna & Slovakia \\
    & 50/200cm Maksutov telescope+Apogee Alta U16M & Crimean Stations of Sternberg  & Crimea \\
    &  & Astronomical Institute (SAI) & \\
    & 250/2000cm+LNI BSI NBI 4k*4k & Caucasian Mountain Observatory & The North Caucasus, Russia \\
    &  & KGO (SAI, KGO) & \\
SRI & CDK 43cm+SBIG STL-1001 & Richard Sabo & Montana, USA \\
Trt & 25cm ALCCD5.2 (QHY6) & Tam\'{a}s Tordai & Budapest, Hungary \\
\hline
\multicolumn{4}{l}{\parbox{500pt}{$^{*}$Observer's code: BSM (Stephen M.~Brincat), COO (Lewis M.~Cook), CRI (Crimean Observatory), GFB (William Goff), IMi (Ian Miller), Ioh (Hiroshi Itoh), Kis (Seiichiro Kiyota), Mhh (Hiroyuki Maehara), Mic (Ra\'{u}l Michel), MLF (Berto Monard), Njh (Kazuhiro Nakajima), OKU (Osaka Kyoiku Univ.~team), 
RPc (Roger D.~Pickard), RIT (Michael Richmond), SGE (Geoff Stone), SHU (Sergey Yu.~ Shugarov, Natalia Katysheva, Polina Golysheva, Olga Vozyakova), SRI (Richard Sabo), Trt (Tam\'{a}s Tordai).}}\\
\multicolumn{4}{l}{$^{\dagger}$itelescope.net.}\\
\multicolumn{4}{l}{$^{\ddagger}$Las Cumbres Observatory Global Telescope Network.}
	\end{tabular}
\end{table*}

\begin{table*}
	\caption{Log of observations of the 2016 outburst of 1SWASP J1621.}
	\label{log}
	\centering
	\begin{tabular}{rrrrrcc}
		\hline
		${\rm Start}^{*}$ & ${\rm End}^{*}$ & ${\rm Mag}^{\dagger}$ 
		& ${\rm Error}^{\ddagger}$ & $N^{\S}$ & ${\rm Obs}^{\parallel}$ & 
		${\rm Band}^{\#}$\\
		\hline
466.9381 & 467.0381 & 14.705 & 0.015 & 47 & Mic & $R_{\rm C}$ \\ 
  466.9387 & 467.0365 & 15.341 & 0.017 & 46 & Mic & $V$ \\ 
  466.9395 & 467.0373 & 16.254 & 0.026 & 46 & Mic & $B$ \\ 
  468.9367 & 468.9931 & 14.845 & 0.027 & 26 & Mic & $R_{\rm C}$ \\ 
  468.9373 & 468.9937 & 15.472 & 0.029 & 26 & Mic & $V$ \\ 
  468.9382 & 468.9922 & 16.348 & 0.029 & 25 & Mic & $B$ \\ 
  505.7052 & 506.0115 & 16.268 & 0.016 & 137 & Mic & $B$ \\ 
  505.7060 & 506.0101 & 14.750 & 0.013 & 136 & Mic & $R_{\rm C}$ \\ 
  505.7066 & 506.0107 & 15.379 & 0.014 & 136 & Mic & $V$ \\ 
  542.8993 & 542.9831 & 13.053 & 0.006 & 38 & Mic & $R_{\rm C}$ \\ 
  542.8999 & 542.9837 & 13.382 & 0.007 & 38 & Mic & $V$ \\ 
  542.9007 & 542.9846 & 13.818 & 0.009 & 38 & Mic & $B$ \\ 
  543.6787 & 543.9624 & 13.501 & 0.014 & 812 & SGE & $CV$ \\ 
  543.8746 & 543.9054 & 14.243 & 0.048 & 34 & HBB & $V$ \\ 
  544.6798 & 544.9507 & 13.804 & 0.020 & 400 & SGE & $CV$ \\ 
  544.7206 & 544.8464 & 13.786 & 0.016 & 313 & COO & $CV$ \\ 
  544.8811 & 544.9792 & 13.519 & 0.052 & 44 & Mic & $R_{\rm C}$ \\ 
  544.8817 & 544.9776 & 13.925 & 0.062 & 43 & Mic & $V$ \\ 
  544.8826 & 544.9784 & 14.449 & 0.079 & 43 & Mic & $B$ \\ 
  545.3104 & 545.5287 & 1.759 & 0.013 & 485 & CRI & $CV$ \\ 
  545.4019 & 545.5645 & 13.891 & 0.026 & 219 & RPc & $V$ \\ 
  545.4241 & 545.5385 & 13.875 & 0.007 & 66 & SHU & $CV$ \\ 
  545.6793 & 545.9420 & 13.885 & 0.016 & 300 & SRI & $CV$ \\ 
  545.7179 & 545.9854 & 14.102 & 0.022 & 362 & SGE & $CV$ \\ 
  545.7365 & 545.9182 & 13.489 & 0.010 & 877 & COO & $R_{\rm C}$ \\ 
  546.3322 & 546.4039 & 14.259 & 0.041 & 101 & Trt & $V$ \\ 
  546.4068 & 546.6473 & 13.621 & 0.023 & 150 & Mic & $R_{\rm C}$ \\ 
  546.4071 & 546.6477 & 14.056 & 0.027 & 150 & Mic & $V$ \\ 
  546.4078 & 546.6483 & 14.624 & 0.038 & 150 & Mic & $B$ \\ 
  546.4359 & 546.5066 & 13.781 & 0.009 & 47 & SHU & $R_{\rm C}$ \\ 
  546.4375 & 546.4540 & 14.009 & 0.008 & 2 & SHU & $CV$ \\ 
  546.4691 & 546.4796 & 14.396 & 0.013 & 14 & JSJ & $B$ \\ 
  546.6157 & 546.7485 & 13.847 & 0.004 & 242 & LCO & $CV$ \\ 
  546.6840 & 546.9420 & 14.179 & 0.017 & 270 & SRI & $CV$ \\ 
  547.2813 & 547.4122 & 2.134 & 0.025 & 31 & CRI & $R_{\rm C}$ \\ 
  547.2853 & 547.4100 & 1.592 & 0.019 & 31 & CRI & $B$ \\ 
  547.2940 & 547.4115 & 1.980 & 0.023 & 30 & CRI & $V$ \\ 
  547.2953 & 547.4128 & 2.229 & 0.025 & 30 & CRI & $I_{\rm C}$ \\ 
  547.3109 & 547.5321 & 2.162 & 0.013 & 500 & CRI & $CV$ \\ 
  547.3536 & 547.4979 & 14.165 & 0.028 & 134 & SHU & $R_{\rm C}$ \\ 
  547.3596 & 547.4988 & 14.220 & 0.019 & 6 & SHU & $CV$ \\ 
  547.4130 & 547.6069 & 14.341 & 0.017 & 351 & IMi & $V$ \\ 
  547.4426 & 547.5714 & 14.179 & 0.005 & 181 & Trt & $V$ \\ 
  547.5081 & 547.6157 & 14.139 & 0.007 & 125 & BSM & $CV$ \\ 
  547.6819 & 547.9724 & 14.355 & 0.012 & 648 & SGE & $CV$ \\ 
  548.2983 & 548.5226 & 2.155 & 0.056 & 52 & CRI & $B$ \\ 
  548.2998 & 548.5200 & 2.439 & 0.040 & 52 & CRI & $V$ \\ 
\hline
\end{tabular}
\end{table*}

\addtocounter{table}{-1}
\begin{table*}
	\caption{Log of observations of the 2016 outburst of 1SWASP J1621 (continued).}
	\label{log}
	\centering
	\begin{tabular}{rrrrrcc}
		\hline
		${\rm Start}^{*}$ & ${\rm End}^{*}$ & ${\rm Mag}^{\dagger}$ 
		& ${\rm Error}^{\ddagger}$ & $N^{\S}$ & ${\rm Obs}^{\parallel}$ & 
		${\rm Band}^{\#}$\\
		\hline
  548.3005 & 548.5207 & 2.526 & 0.032 & 52 & CRI & $R_{\rm C}$ \\ 
  548.3011 & 548.5213 & 2.557 & 0.025 & 52 & CRI & $I_{\rm C}$ \\ 
  548.3439 & 548.4874 & 14.802 & 0.027 & 93 & SHU, NKa & $R_{\rm C}$ \\ 
  548.3584 & 548.5426 & 2.473 & 0.012 & 416 & CRI & $CV$ \\ 
  548.3680 & 548.4865 & 14.411 & 0.043 & 4 & SHU & $CV$ \\ 
  548.3890 & 548.6118 & 14.598 & 0.014 & 266 & BPO & $CV$ \\ 
  548.4251 & 548.5936 & 14.716 & 0.016 & 314 & IMi & $V$ \\ 
  548.5078 & 548.6157 & 14.456 & 0.010 & 119 & BSM & $CV$ \\ 
  548.6831 & 548.9865 & 14.714 & 0.012 & 541 & SGE & $CV$ \\ 
  549.3060 & 549.4802 & 2.406 & 0.016 & 44 & CRI & $B$ \\ 
  549.3074 & 549.4776 & 2.668 & 0.016 & 43 & CRI & $V$ \\ 
  549.3082 & 549.4783 & 2.728 & 0.017 & 43 & CRI & $R_{\rm C}$ \\ 
  549.3087 & 549.4789 & 2.699 & 0.017 & 43 & CRI & $I_{\rm C}$ \\ 
  549.6008 & 549.8552 & 15.795 & 0.031 & 95 & CMJ & $B$ \\ 
  549.6020 & 549.8564 & 15.010 & 0.020 & 96 & CMJ & $V$ \\ 
  549.6277 & 549.8538 & 14.966 & 0.014 & 176 & RIT & $CV$ \\ 
  549.6900 & 549.9682 & 14.985 & 0.010 & 548 & SGE & $CV$ \\ 
  549.6976 & 549.9180 & 15.741 & 0.038 & 89 & GFB & $B$ \\ 
  549.6985 & 549.9190 & 14.985 & 0.023 & 90 & GFB & $V$ \\ 
  550.0408 & 550.1700 & 14.838 & 0.010 & 338 & OKU & $CV$ \\ 
  550.0501 & 550.2506 & 14.832 & 0.012 & 308 & Kis & $CV$ \\ 
  550.0555 & 550.2780 & 14.625 & 0.010 & 284 & Ioh & $R_{\rm C}$ \\ 
  550.2851 & 550.5242 & 2.734 & 0.030 & 60 & CRI & $B$ \\ 
  550.2866 & 550.5216 & 2.921 & 0.023 & 59 & CRI & $V$ \\ 
  550.2873 & 550.5223 & 2.936 & 0.021 & 59 & CRI & $R_{\rm C}$ \\ 
  550.2879 & 550.5229 & 2.860 & 0.018 & 59 & CRI & $I_{\rm C}$ \\ 
  550.3179 & 550.6242 & 14.940 & 0.012 & 183 & BSM & $CV$ \\ 
  550.3451 & 550.5412 & 15.016 & 0.082 & 6 & SHU & $CV$ \\ 
  550.6839 & 550.9838 & 15.127 & 0.009 & 541 & SGE & $CV$ \\ 
  550.6942 & 550.9266 & 15.103 & 0.015 & 136 & SRI & $CV$ \\ 
  550.6974 & 550.9731 & 15.972 & 0.020 & 230 & GFB & $B$ \\ 
  551.2732 & 551.5204 & 2.999 & 0.028 & 62 & CRI & $B$ \\ 
  551.2747 & 551.5218 & 3.096 & 0.025 & 62 & CRI & $V$ \\ 
  551.2754 & 551.5226 & 3.061 & 0.022 & 62 & CRI & $R_{\rm C}$ \\ 
  551.2760 & 551.5232 & 2.963 & 0.019 & 62 & CRI & $I_{\rm C}$ \\ 
  552.4002 & 552.4594 & 15.087 & 0.021 & 32 & SHU, Nka & $CV$ \\ 
  552.6653 & 552.8563 & 16.368 & 0.024 & 72 & CMJ & $B$ \\ 
\hline
\end{tabular}
\end{table*}

\addtocounter{table}{-1}
\begin{table*}
	\caption{Log of observations of the 2016 outburst of 1SWASP J1621 (continued).}
	\label{log}
	\centering
	\begin{tabular}{rrrrrcc}
		\hline
		${\rm Start}^{*}$ & ${\rm End}^{*}$ & ${\rm Mag}^{\dagger}$ 
		& ${\rm Error}^{\ddagger}$ & $N^{\S}$ & ${\rm Obs}^{\parallel}$ & 
		${\rm Band}^{\#}$\\
		\hline
  552.6666 & 552.8550 & 15.385 & 0.023 & 70 & CMJ & $V$ \\ 
  553.5893 & 553.8566 & 16.317 & 0.024 & 103 & CMJ & $B$ \\ 
  553.5905 & 553.8553 & 15.325 & 0.017 & 102 & CMJ & $V$ \\ 
  553.6977 & 553.9739 & 16.306 & 0.014 & 230 & GFB & $B$ \\ 
  553.7135 & 553.8156 & 15.181 & 0.012 & 26 & SGE & $V$ \\ 
  554.0120 & 554.1428 & 15.103 & 0.007 & 289 & OKU & $CV$ \\ 
  554.3890 & 554.5240 & 15.312 & 0.018 & 71 & PVE & $V$ \\ 
  554.4558 & 554.5183 & 15.092 & 0.012 & 67 & SHU, Nka & $CV$ \\ 
  554.6905 & 554.9057 & 15.333 & 0.024 & 72 & SGE & $V$ \\ 
  554.6935 & 554.9087 & 16.363 & 0.036 & 72 & SGE & $B$ \\ 
  555.3449 & 555.5094 & 14.169 & 0.066 & 9 & PVE & $I_{\rm C}$ \\ 
  555.3524 & 555.5439 & 15.359 & 0.025 & 55 & PVE & $V$ \\ 
  555.3917 & 555.5158 & 16.326 & 0.085 & 8 & PVE & $B$ \\ 
  555.6884 & 555.9814 & 15.211 & 0.023 & 49 & SGE & $V$ \\ 
  555.6913 & 555.9845 & 16.242 & 0.028 & 53 & SGE & $B$ \\ 
  556.3195 & 556.5463 & 14.141 & 0.019 & 61 & PVE & $I_{\rm C}$ \\ 
  556.3574 & 556.5290 & 16.347 & 0.085 & 8 & PVE & $B$ \\ 
  556.3631 & 556.5347 & 15.311 & 0.050 & 8 & PVE & $V$ \\ 
  556.6001 & 556.8552 & 16.352 & 0.021 & 99 & CMJ & $B$ \\ 
  556.6014 & 556.8539 & 15.352 & 0.018 & 97 & CMJ & $V$ \\ 
  556.6899 & 556.9798 & 15.297 & 0.020 & 94 & SGE & $V$ \\ 
  556.6930 & 556.9829 & 16.310 & 0.026 & 94 & SGE & $B$ \\ 
  557.3278 & 557.4053 & 14.093 & 0.034 & 14 & PVE & $I_{\rm C}$ \\ 
  557.3304 & 557.4133 & 15.278 & 0.045 & 15 & PVE & $V$ \\ 
  558.3199 & 558.5466 & 15.373 & 0.019 & 99 & PVE & $V$ \\ 
  563.0314 & 563.0755 & 14.638 & 0.010 & 62 & OKU & $R_{\rm C}$ \\ 
  566.0187 & 566.1714 & 14.774 & 0.013 & 160 & OKU & $R_{\rm C}$ \\ 
  617.3656 & 617.5738 & 15.229 & 0.009 & 288 & RPc & $CV$ \\
  630.3350 & 630.5050 & 15.238 & 0.009 & 236 & RPc & $CV$ \\
  665.2928 & 665.3731 & 15.202 & 0.012 & 106 & RPc & $CV$ \\
  823.4359 & 823.5977 & 14.696 & 0.143 & 30 & WTH & $R_{\rm C}$ \\ 
\hline
\multicolumn{7}{l}{$^{*}$BJD--2457000.0.}\\
\multicolumn{7}{l}{$^{\dagger}$Mean magnitude.  Here, CRI reports the relative magnitude.}\\
\multicolumn{7}{l}{$^{\ddagger}1\sigma$ of mean magnitude.}\\
\multicolumn{7}{l}{$^{\S}$Number of observations.}\\
\multicolumn{7}{l}{\parbox{270pt}{$^{\parallel}$Observer's code: GFB (William Goff), Kis (Seiichiro Kiyota), COO (Lewis M.~Cook), Ioh (Hiroshi Itoh), CRI (Crimean Observatory), OKU (Osaka Kyoiku Univ.~team), Trt (Tam\'{a}s Tordai), RPc (Roger D.~Pickard), IMi (Ian Miller), RIT (Michael Richmond), SGE (Geoff Stone), SHU (S.~Shugarov team), SRI (Richard Sabo), LCO (Colin Littlefield), NKa (Natalia Katysheva \& Sergei Yu.~Shugarov), PVE (Velimir Popov), HBB (Barbara Harris), CMJ (Michael Cook), WTH (Wikander, Thomas), Mic (Ra\'{u}l Michel), BSM (Stephen M.~Brincat), JSJ (Steve Johnson).  }}\\
\multicolumn{7}{l}{\parbox{240pt}{$^{\#}$Filter. ``$CV$'' means no (clear) filter.}}\\
\end{tabular}
\end{table*}

\begin{table*}
	\caption{Log of observations of the 2006 outburst and the 2013 quiescence of BD Pav.}
	\label{logbdpav}
	\centering
	\begin{tabular}{rrrrrcc}
		\hline
		${\rm Start}^{*}$ & ${\rm End}^{*}$ & ${\rm Mag}^{\dagger}$ 
		& ${\rm Error}^{\ddagger}$ & $N^{\S}$ & ${\rm Obs}^{\parallel}$ & 
		${\rm Band}^{\#}$\\  
		\hline
  979.2944 & 79.4928 & 12.481 & 0.003 & 539 & MLF & $CV$ \\ 
  980.1966 & 80.5440 & 12.540 & 0.002 & 903 & MLF & $CV$ \\ 
  982.1992 & 82.5138 & 12.635 & 0.003 & 851 & MLF & $CV$ \\ 
  983.2130 & 83.5227 & 12.699 & 0.003 & 880 & MLF & $CV$ \\ 
  984.1914 & 84.5195 & 12.776 & 0.003 & 930 & MLF & $CV$ \\ 
  985.1892 & 85.5080 & 12.965 & 0.003 & 858 & MLF & $CV$ \\ 
  3454.7143 & 2554.8467 & 15.152 & 0.114 & 150 & OAR & $V$ \\ 
  3455.6887 & 2555.9171 & 15.157 & 0.144 & 220 & OAR & $V$ \\ 
  3457.7425 & 2557.9143 & 15.101 & 0.109 & 167 & OAR & $V$ \\ 
  3460.7053 & 2560.8474 & 15.039 & 0.030 & 9 & OAR & $V$ \\ 
  3461.7647 & 2561.7685 & 14.956 & 0.015 & 5 & OAR & $V$ \\ 
  3462.7649 & 2562.7688 & 15.077 & 0.033 & 5 & OAR & $V$ \\ 
  3463.7621 & 2563.7657 & 15.156 & 0.016 & 5 & OAR & $V$ \\ 
  \hline
\multicolumn{7}{l}{$^{*}$BJD--2453000.0.}\\
\multicolumn{7}{l}{$^{\dagger}$Mean magnitude.}\\
\multicolumn{7}{l}{$^{\ddagger}1\sigma$ of mean magnitude.}\\
\multicolumn{7}{l}{$^{\S}$Number of observations.}\\
\multicolumn{7}{l}{\parbox{240pt}{$^{\parallel}$Observer's code: MLF (Berto Monard).}}\\
\multicolumn{7}{l}{\parbox{240pt}{$^{\#}$Filter.  ``$CV$'' means no (clear) filter.}}\\
\end{tabular}
\end{table*}

\begin{table*}
	\caption{Log of observations of the 2009 outburst of V364 Lib.}
	\label{logV364 Lib}
	\centering
	\begin{tabular}{rrrrrcc}
		\hline
		${\rm Start}^{*}$ & ${\rm End}^{*}$ & ${\rm Mag}^{\dagger}$ 
		& ${\rm Error}^{\ddagger}$ & $N^{\S}$ & ${\rm Obs}^{\parallel}$ & 
		${\rm Band}^{\#}$\\  
		\hline
928.0757 & 28.2011 & 10.412 & 0.001 & 405 & Mhh & $V$ \\ 
  928.1202 & 28.2177 & 11.478 & 0.003 & 223 & Njh & $CV$ \\ 
  928.2010 & 28.2883 & 1.822 & 0.001 & 481 & OUS & $CV$ \\ 
  928.2017 & 28.2036 & 10.569 & 0.006 & 6 & Mhh & $B$ \\ 
  929.1080 & 29.1733 & 10.711 & 0.001 & 89 & Kis & $CV$ \\ 
  929.1197 & 29.2284 & 1.893 & 0.002 & 351 & OUS & $CV$ \\ 
  930.0970 & 30.2722 & 11.460 & 0.001 & 412 & Njh & $CV$ \\ 
  930.1155 & 30.2758 & 1.880 & 0.001 & 885 & OUS & $CV$ \\ 
  930.1287 & 30.1968 & 10.732 & 0.002 & 96 & Kis & $CV$ \\ 
  930.1431 & 30.2359 & 10.436 & 0.002 & 138 & Mhh & $V$ \\ 
  930.1434 & 30.2355 & 10.628 & 0.002 & 136 & Mhh & $B$ \\ 
  931.0911 & 31.3198 & 11.493 & 0.001 & 537 & Njh & $CV$ \\ 
  931.0944 & 31.1718 & 10.710 & 0.003 & 104 & Kis & $CV$ \\ 
  932.0938 & 32.2945 & 11.552 & 0.001 & 465 & Njh & $CV$ \\ 
  932.1091 & 32.1847 & 10.770 & 0.003 & 102 & Kis & $CV$ \\ 
  932.1812 & 32.2890 & 10.729 & 0.001 & 410 & Mhh & $B$ \\ 
  937.1064 & 37.1435 & 11.750 & 0.002 & 88 & Njh & $CV$ \\ 
  938.0842 & 38.2074 & 11.906 & 0.003 & 237 & Njh & $CV$ \\ 
  939.2176 & 39.3233 & 11.199 & 0.001 & 591 & Mhh & $CV$ \\ 
  945.1195 & 45.2613 & 12.179 & 0.001 & 335 & Njh & $CV$ \\ 
  947.1338 & 47.1456 & 12.247 & 0.005 & 29 & Njh & $CV$ \\ 
  949.0912 & 49.2735 & 12.347 & 0.002 & 422 & Njh & $CV$ \\ 
  950.1414 & 50.2637 & 12.363 & 0.002 & 288 & Njh & $CV$ \\ 
  951.0445 & 51.2736 & 12.407 & 0.002 & 540 & Njh & $CV$ \\ 
  951.3060 & 51.6199 & 0.042 & 0.001 & 1623 & MLF & $V$ \\ 
  952.0562 & 52.1361 & 12.431 & 0.002 & 189 & Njh & $CV$ \\ 
  953.0907 & 53.2577 & 12.433 & 0.001 & 392 & Njh & $CV$ \\ 
  965.0846 & 65.2432 & 12.440 & 0.001 & 370 & Njh & $CV$ \\ 
   \hline
\multicolumn{7}{l}{$^{*}$BJD--2454000.0.}\\
\multicolumn{7}{l}{$^{\dagger}$Mean magnitude.  Here, OUS reports the relative magnitude.  }\\
\multicolumn{7}{l}{$^{\ddagger}1\sigma$ of mean magnitude.}\\
\multicolumn{7}{l}{$^{\S}$Number of observations.}\\
\multicolumn{7}{l}{\parbox{240pt}{$^{\parallel}$Observer's code: Mhh (Hiroyuki Maehara), Njh (Kazuhiro Nakajima), OUS (Okayama U.~of Science), Kis (Seiichiro Kiyota).}}\\
\multicolumn{7}{l}{\parbox{240pt}{$^{\#}$Filter.  ``$CV$'' means no (clear) filter.}}\\
\end{tabular}
\end{table*}

\begin{table*}
	\caption{Log of spectroscopic observations of V364 Lib.}
	\label{logV364Libspe}
	\centering
	\begin{tabular}{ccrrrccc}
		\hline
		Date$^{*}$ & Name$^{\dagger}$ & Start Time$^{\ddagger}$ & End Time$^{\S}$ & Exp$^{\parallel}$ & Number$^{\#}$ & Grating$^{\P}$ & \textcolor{black}{Site}$^{**}$\\  
		\hline
2009-04-07 & V364 Lib & 54929.1854 & 54929.2062 & 60 & 11 & Grism & G \\ 
2009-04-09 & V364 Lib & 54931.1862 & 54931.2022 & 60 & 15 & Grism & G \\ 
2009-04-10 & V364 Lib & 54932.1536 & 54932.1703 & 60 & 12 & Grism & G \\ 
 -- & V364 Lib & 54932.1953 & 54932.2043 & 60 & 9 & Grism & G \\ 
2009-04-15 & V364 Lib & 54937.0754 & 54937.0764 & 30 & 3 & $I_{\rm C}$ & G \\ 
 -- & V364 Lib & 54937.0816 & 54937.0948 & 60 & 11 & Grism & G \\ 
2009-04-18 & V364 Lib & 54940.1137 & 54940.1151 & 60 & 2 & Grism & G \\ 
2009-04-19 & V364 Lib & 54941.0797 & 54941.0901 & 60 & 10 & Grism & G \\ 
2009-04-22 & V364 Lib & 54944.0798 & 54944.0902 & 60 & 10 & Grism & G \\ 
2009-04-23 & V364 Lib & 54945.1173 & 5494.1257 & 60 & 5 & Grism & G \\ 
2009-04-26 & V364 Lib & 54948.1084 & 54948.1209 & 120 & 5 & Grism & G \\ 
2009-04-27 & V364 Lib & 54949.0751 & 54949.0820 & 60 & 6 & Grism & G \\ 
2009-04-28 & V364 Lib & 54950.1300 & 54950.1473 & 60 & 12 & Grism & G \\ 
2009-05-02 & V364 Lib & 54954.0627 & 54954.0738 & 60 & 10 & Grism & G \\ 
2009-05-05 & V364 Lib & 54961.8932 & 54961.9392 & 900 & 5 & std-Yd & S \\
-- & V364 Lib & 54961.9619 & 54961.0183 & 900 & 6 & std-Bc & S \\
-- & V364 Lib & 54962.0348 & 54962.0573 & 900 & 3 & std-Yd & S \\
2009-05-08 & V364 Lib & 54960.1634 & 54960.0759 & 60 & 10 & Grism & G \\  
2009-05-13 & V364 Lib & 54965.1412 & 54965.0419 & 20 & 3 & R & G \\ 
 -- & V364 Lib & 54965.1510 & 54965.1579 & 60 & 7 & Grism & G \\ 
2009-05-20 & V364 Lib & 54972.1113 & 54972.1183 & 60 & 6 & Grism & G \\ 
\hline
\multicolumn{8}{l}{$^{*}$Observational date (Japan Standard Time).}\\
\multicolumn{8}{l}{$^{\dagger}$Name of star.  }\\
\multicolumn{8}{l}{$^{\ddagger}$Start time of observations in the unit of BJD--2400000.}\\
\multicolumn{8}{l}{$^{\S}$End time of observations in the unit of BJD--2400000.}\\
\multicolumn{8}{l}{$^{\parallel}$Exposure time of each observation in unit of seconds.}\\
\multicolumn{8}{l}{$^{\#}$Number of observations.}\\
\multicolumn{8}{l}{$^{\P}$Diffraction or echelle gratings.  }\\
\multicolumn{8}{l}{$^{**}$Name of observatory: S (Subaru Observatory), G (Gunma Observatory).  }\\
\end{tabular}
\end{table*}

\begin{table*}
\caption{Radial velocity measured in the 2009 quiescence of V364 Lib.}
\centering
\begin{tabular}{rrrr}
  \hline
Time$^{*}$ & Phase$^{\dagger}$ & Radial Velocity & Error \\ 
  \hline
  54929.1934 & 0.7912 & $-$107.8 & 10.8 \\ 
  54931.1895 & 0.6329 & $-$50.3 & 11.3 \\ 
  54932.1952 & 0.0648 & 73.3 & 13.7 \\ 
  54937.0838 & 0.0243 & 42.0 & 22.2 \\ 
  54941.0801 & 0.7135 & $-$99.8 & 37.9 \\ 
  54944.1177 & 0.0379 & $-$5.6 & 46.4 \\ 
  54945.1577 & 0.5186 & $-$12.3 & 52.5 \\ 
   \hline
\multicolumn{4}{l}{$^{*}$Time of observations in the unit of BJD--2400000.}\\
\multicolumn{4}{l}{\parbox{190pt}{$^{\dagger}$Orbital phase under the assumption that the orbital period is 0.7024293 d.  }}\\   
\end{tabular}
\end{table*}

\begin{table*}
\caption{Radial velocity measured in the 2009 outburst of V364 Lib.}
\centering
\begin{tabular}{rrrr}
  \hline
Time$^{*}$ & Phase$^{\dagger}$ & Radial Velocity & Error \\  
  \hline
  54961.3873 & 0.3746 & $-$49.7 & 4.3 \\ 
  54961.3994 & 0.3917 & $-$38.1 & 3.3 \\ 
  54961.4107 & 0.4078 & $-$34.3 & 3.7 \\ 
  54961.4220 & 0.4239 & $-$28.4 & 3.7 \\ 
  54961.4334 & 0.4401 & $-$21.1 & 4.1 \\ 
  54961.4561 & 0.4724 & $-$6.0 & 5.3 \\ 
  54961.4676 & 0.4888 & 2.4 & 4.3 \\ 
  54961.4788 & 0.5048 & 8.5 & 4.2 \\ 
  54961.4901 & 0.5208 & 13.4 & 3.1 \\ 
  54961.5013 & 0.5367 & 22.5 & 3.1 \\ 
  54961.5125 & 0.5527 & 30.8 & 2.2 \\ 
  54961.5289 & 0.5762 & 40.5 & 6.7 \\ 
  54961.5403 & 0.5923 & 49.5 & 5.5 \\ 
  54961.5515 & 0.6083 & 51.4 & 4.5 \\ 
   \hline
\multicolumn{4}{l}{$^{*}$Time of observations in the unit of BJD--2400000.}\\
\multicolumn{4}{l}{\parbox{190pt}{$^{\dagger}$Orbital phase under the assumption that the orbital period is 0.7024293 d.  }}\\      
\end{tabular}
\end{table*}

\begin{figure*}[htb]
\addtocounter{figure}{-18}
\vspace{-3mm}
\begin{minipage}{1.0\hsize}
\begin{center}
\FigureFile(145mm, 50mm){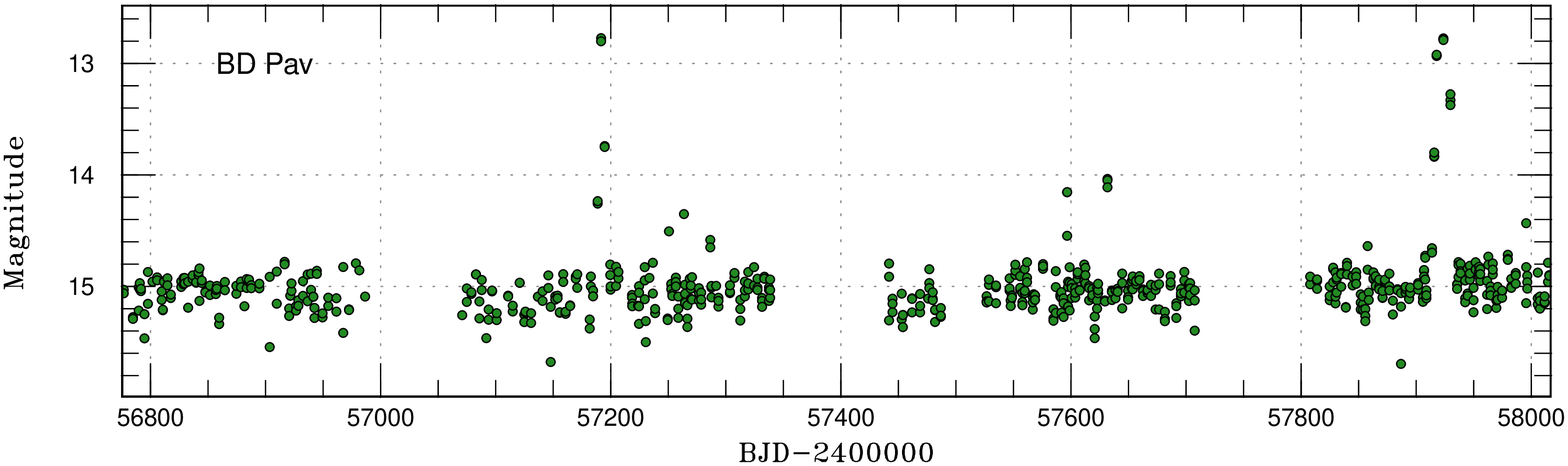}
\end{center}
\end{minipage}
\\
\begin{minipage}{1.0\hsize}
\begin{center}
\FigureFile(145mm, 50mm){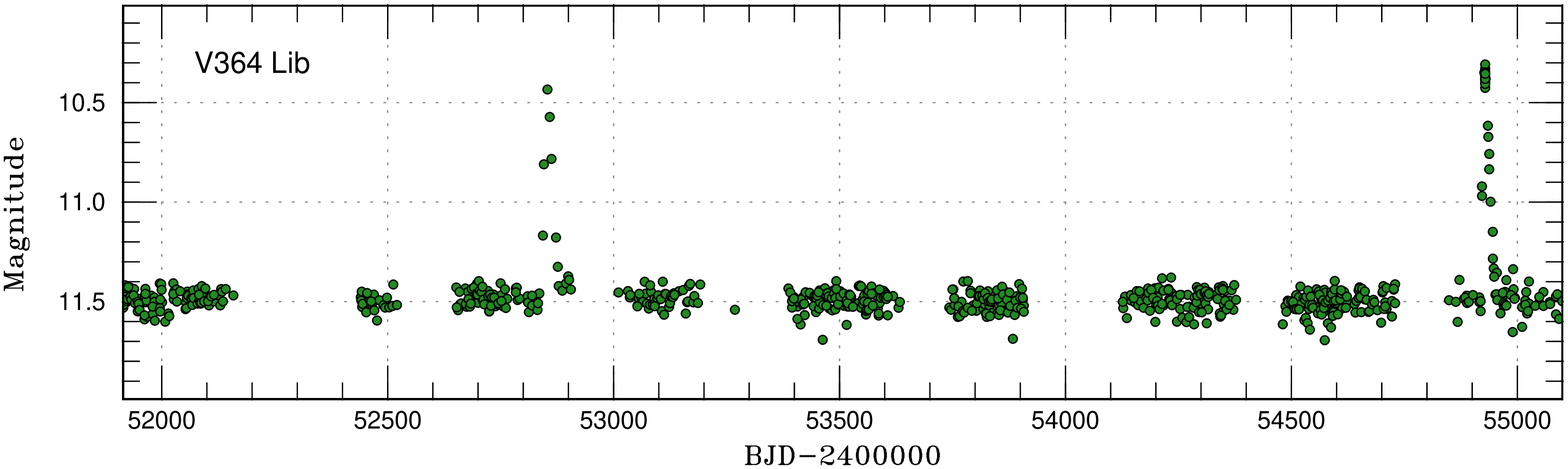}
\end{center}
\end{minipage}
\\
\begin{minipage}{1.0\hsize}
\begin{center}
\FigureFile(145mm, 50mm){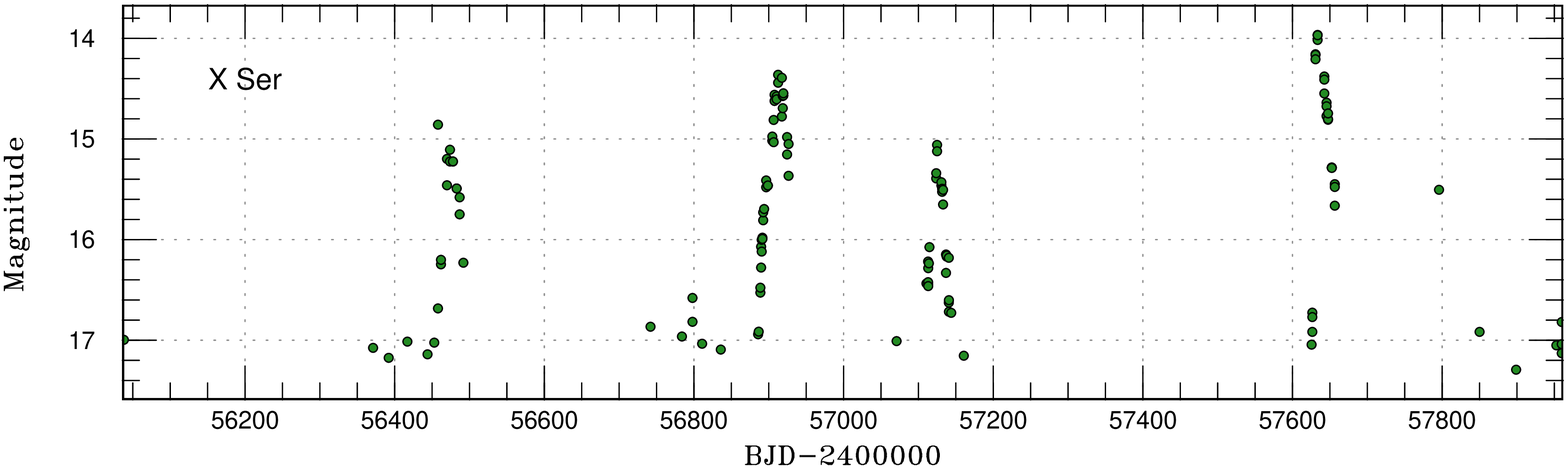}
\end{center}
\end{minipage}
\\
\begin{minipage}{1.0\hsize}
\begin{center}
\FigureFile(145mm, 50mm){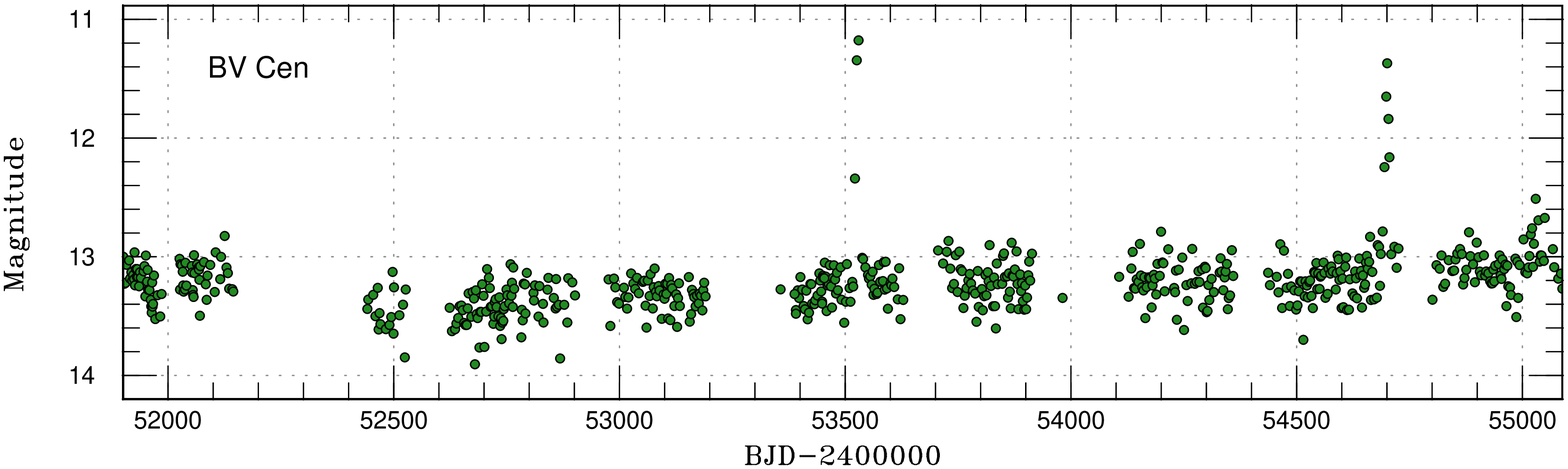}
\end{center}
\end{minipage}
\\
\begin{minipage}{1.0\hsize}
\begin{center}
\FigureFile(145mm, 50mm){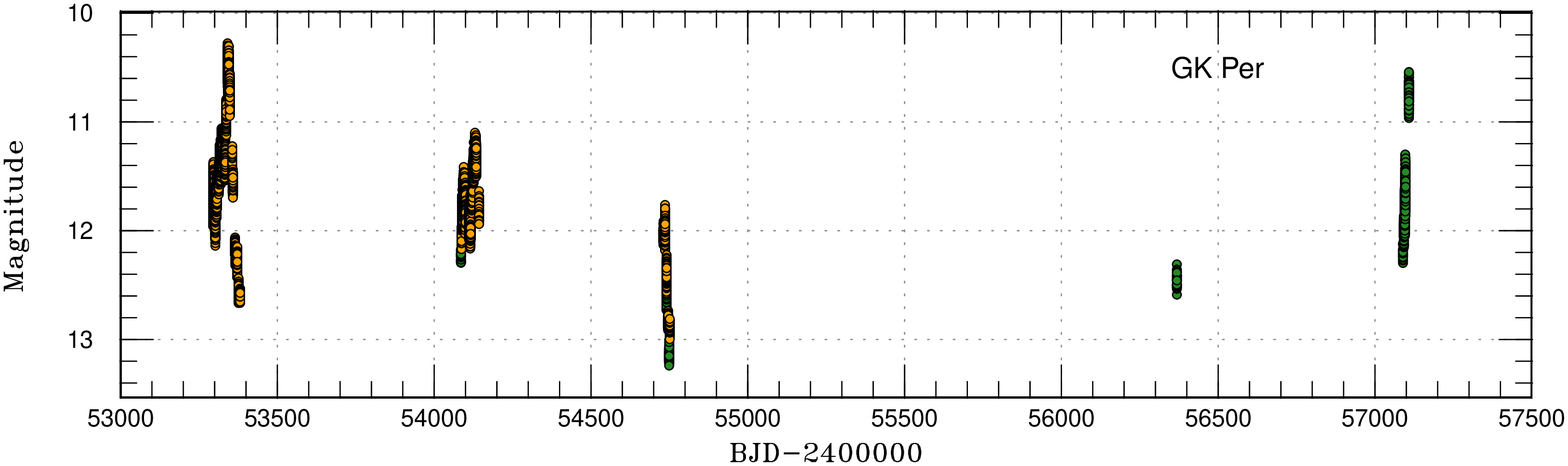}
\vspace{3mm}
\end{center}
\end{minipage}
\caption{Long-term light curves in the long-period objects discussed in this paper.  The light curves of V364 Lib and BV Cen are derived from the ASAS-3 data archive.  The light curves of BD Pav and X Ser are derived from the ASAS-SN data archive.  Green and orange points represent the observations in the $V$ band and no (clear) filter, respectively.}
\label{longtermlc}
\end{figure*}

\end{document}